\documentclass[a4paper,11pt]{article} %11pt in original
\pdfoutput=1 % if your are submitting a pdflatex (i.e. if you have
\usepackage{jheppub}
\usepackage[whole]{bxcjkjatype}  %%日本語環境

\usepackage{subfig}
\usepackage{amsmath, amssymb}
\usepackage{mathrsfs}
\usepackage{graphicx}% Include figure files
\usepackage{dcolumn}% Align table columns on decimal point
\usepackage{bm}% bold math
\renewcommand{\labelenumi}{(\roman{enumi})}
\usepackage{comment}
\usepackage{braket}
\usepackage{appendix}
\usepackage{xcolor}
\usepackage{amsthm}
\usepackage{bbm}
\usepackage{bbm,bm,graphicx,mathtools,color,hyperref,slashed}

\newcommand{\diff}{\mathrm{d}}

\newcommand{\Diff}{{\mathcal{D}}}

\newcommand{\im}{\mathrm{i}}

\newcommand{\rme}{\mathrm{e}}

\definecolor{darkred}{rgb}{0.7, 0.0, 0.0}
\newcommand\hl[1]{{\color{black} #1}}
\newcommand{\gauge}{\mathrm{gauge}}

\title{Semiclassics for the QCD vacuum structure through $T^2$-compactification with the baryon-'t Hooft flux}

\author[]{Yui Hayashi}
\emailAdd{yui.hayashi@yukawa.kyoto-u.ac.jp}

\author[]{and Yuya Tanizaki}
\emailAdd{yuya.tanizaki@yukawa.kyoto-u.ac.jp}

\affiliation[]{Yukawa Institute for Theoretical Physics, Kyoto University, Kitashirakawa Oiwakecho, Sakyo-ku, Kyoto 606-8502, Japan
}

\preprint{YITP-24-15}

\abstract{
We study QCD vacuum structure with the topological $\theta$ angle using a recently proposed semiclassical approach on $\mathbb{R}^2 \times T^2$ with the 't Hooft and baryon magnetic fluxes.
Under the assumption of adiabatic continuity in this setup, the confining vacuum can be described by the dilute gas of center vortices.
With this semiclassical approach, we derive the 2d effective description at small $T^2$ and successfully explain the reasonable theta dependence of the QCD vacuum: In the one-flavor QCD at $\theta = \pi$, the $CP$ symmetry is spontaneously broken for quark mass above a critical value and restored for a subcritical mass, while the $CP$ symmetry is always spontaneously broken in the multi-flavor QCD at $\theta = \pi$.
From our semiclassical description, we discuss implications to the $4$d chiral Lagrangian and propose how the $\eta'$ meson should be incorporated in consistent with known global structures: The periodicity of the $\eta'$ should be extended from the naive one $2\pi$ to $2\pi N$. 
Additionally, we revisit the phase diagram of $N_f = 1+1$ and $N_f = 1+1+1$ QCD on the up and down quark mass plane, confirming and refining the existence of the $CP$-broken Dashen phase.
}

\begin{document}

\maketitle

\section{Introduction}

Understanding quark confinement \cite{Wilson:1974sk} is a fundamental and long-standing problem in strong interactions.
As well as the dual superconductor picture, i.e. confinement due to monopole condensation \cite{Nambu:1974zg, Mandelstam:1974pi, Polyakov:1975rs, tHooft:1977nqb, Polyakov:1976fu, Polyakov:1978vu, tHooft:1979rtg, tHooft:1981bkw, Unsal:2007jx}, the center-vortex picture is one of the promising explanations for the mechanism of quark confinement \cite{tHooft:1977nqb, Cornwall:1979hz, Nielsen:1979xu, Ambjorn:1980ms, DelDebbio:1996lih, Faber:1997rp, Kovacs:1998xm, DelDebbio:1998luz, deForcrand:1999our, Langfeld:1998cz, Engelhardt:1999fd, Alexandrou:1999vx, Ambjorn:1999ym, Diakonov:2002bx} (see \cite{Greensite:2011zz, Greensite:2003bk} for an introductory review).
The proliferation of codimension-2 magnetic defects would lead to the area-law falloff of the Wilson loop average by randomly rotating the phase of the Wilson loop, and the 1-form center symmetry is restored in the modern terminology. 

As a related topic, the Yang-Mills/QCD vacuum structure with $\theta$ angle has been intensively studied for decades \cite{Dashen:1970et, Callan:1976je, Jackiw:1976pf, tHooft:1976rip, Coleman:1976uz, Witten:1979ey, Witten:1979vv, Rosenzweig:1979ay, Nath:1979ik, Witten:1980sp, DiVecchia:1980yfw, Kawarabayashi:1980dp, tHooft:1981bkw, Ohta:1981ai, Cardy:1981fd, Cardy:1981qy, Wiese:1988qz,Affleck:1991tj, Creutz:1995wf, Smilga:1998dh, Witten:1998uka, Halperin:1998rc, Tytgat:1999yx,Creutz:2003xc,Creutz:2003xu, Boer:2008ct, Boomsma:2009eh, Creutz:2009kx, DElia:2012pvq, DElia:2013uaf, Creutz:2013xfa, Aoki:2014moa, Mameda:2014cxa}.
Recently, utilizing generalized global symmetries and their 't Hooft anomalies has attracted interests as a prevalent method for investigating the vacuum structure of QCD and QCD-like theories (for an incomplete list, see \cite{tHooft:1979rat, Wen:2013oza, Kapustin:2014lwa, Kapustin:2014gua,Gaiotto:2014kfa, Gaiotto:2017yup, Gaiotto:2017tne, Kikuchi:2017pcp, Shimizu:2017asf, Komargodski:2017dmc, Komargodski:2017smk, Tanizaki:2017bam, Tanizaki:2017mtm, Tanizaki:2017qhf, Sulejmanpasic:2018upi, Tanizaki:2018wtg, Tanizaki:2018xto, Karasik:2019bxn, Anber:2018jdf, Anber:2018tcj, Anber:2019nze, Yonekura:2019vyz, Sulejmanpasic:2020zfs}). 
The vacuum structure is intimately associated with the confinement mechanism.
For example, the (intuitive) dual superconductor picture roughly explains the expected $\theta$-dependence of the confining vacua.
By the Witten effect, if the vacuum at $\theta = 0$ is the monopole-condensed phase, the dyon-condensed phase appears at $\theta = 2\pi$, and there would be a level crossing at $\theta = \pi$.
Studying the response to the topological angle would be a significant subject to cultivate our intuition on the confining vacuum.

\begin{figure}[t]
\centering
\includegraphics[scale=0.6]{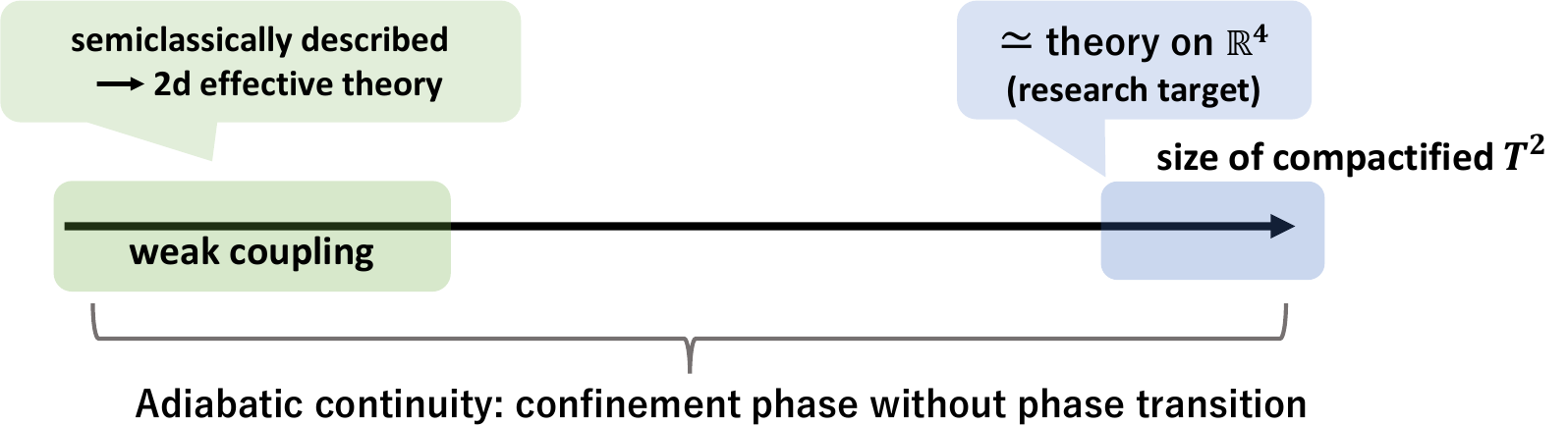
}
\caption{The main ansatz for our semiclassical approach, the adiabatic continuity conjecture.
The adiabatic continuity conjecture claims that no phase transition happens as the torus size changes for the 't Hooft flux $T^2$ compactification.
We try to investigate confining vacua of the strongly-coupled theory on $\mathbb{R}^4$ from a weakly-coupled theory on $\mathbb{R}^2 \times T^2$ at a small torus $T^2$.
}
\label{fig:methodology}
\end{figure}

\hl{Recently, a novel semiclassical approach to exploring confining vacua has been proposed~\cite{Tanizaki:2022ngt}.
This semiclassical framework is constructed on $\mathbb{R}^2 \times T^2$ at a small torus $T^2$ with 't Hooft flux \cite{tHooft:1979rtg, tHooft:1981sps} (combined with flavor or $U(1)_B$ twist if necessary), 
which maintains 't~Hooft anomalies of $4$d theories in the $2$d effective theory~\cite{Tanizaki:2017qhf, Yamazaki:2017dra, Tanizaki:2022ngt}. 
This $2$d effective theory gives the weak-coupling description of the confinement-related phenomena, and it successfully offers a reasonable picture for various confining theories \cite{Tanizaki:2022ngt, Tanizaki:2022plm, Hayashi:2023wwi}, including $SU(N)$ Yang-Mills theory, $\mathcal{N} = 1$ super-Yang-Mills theory, QCD with fundamental quarks, QCD with 2-index quarks, and bifundamental QCD.
Based on these observations, let us assume the adiabatic continuity conjecture in this paper: the $2$d semiclassical theory captures qualitative features of $4$d strongly-coupled theories. 
Figure \ref{fig:methodology} provides a summary of this methodology.
}

In this paper, we extend upon the semiclassical analysis for QCD with fundamental quarks. While Ref.~\cite{Tanizaki:2022ngt} already proposed a method and briefly examined its relationship with the chiral Lagrangian as well as its consistency with the Witten-Veneziano formula, the detailed analysis of the vacuum structure has not been carried out (e.g., in the presence of quark mass). 
Our focus here is to expand this analysis, particularly exploring how the proliferation of center vortices elucidates the QCD vacuum structure.

The main aim of this analysis is to clarify and support the well-accepted picture of the QCD vacuum: QCD with a single quark ($N_f = 1$) at $\theta = \pi$ exhibits spontaneous $CP$ breaking for quark mass exceeding a critical value, while the vacuum remains trivial for subcritical quark mass. For $N_f \geq 2$, the $CP$ breaking occurs at $\theta = \pi$ for any quark mass (see Fig.~\ref{fig:multi_flavor_phase_diagram}). 
We will also comment on related topics, such as the significance of the global structure of $\eta'$ meson, subtleties on domain walls, and discrete anomalies.
These discussions aim to provide a more comprehensive understanding of the QCD vacuum structure through the lens of center-vortex proliferation.

As an application, we revisit the phase diagram of $N_f = 1+1$ QCD (and $N_f = 1+1+1$ QCD with a fixed strange mass)\footnote{\hl{In this paper, we use the jargon $N_f = 1+1$ or $(1+1)$-flavor when the up and down quarks have different masses. 
Similarly, $N_f = 1+1+1$ QCD indicates that up, down, and strange masses are all different.}} on the $(m_u, m_d)$ plane, examining it through our semiclassical method.
The physics of $N_f = 1+1$ and $N_f = 1+1+1$ QCD on the $(m_u, m_d)$ plane has gathered interest to deepen our understanding of chiral symmetry breaking and also in the context of the strong $CP$ problem. 
An interesting phenomenon in this study is the appearance of a spontaneously $CP$-broken phase, known as the Dashen phase~\cite{Dashen:1970et}. 
This phase is expected to appear in certain regions where $m_u m_d < 0$ with $\theta=0$, which is equivalent to $\theta=\pi$ with both $m_u, m_d>0$.

Previous investigations into this topic basically use the phenomenological chiral Lagrangian, both with and without an $\eta'$ contribution~\cite{Creutz:1995wf, Creutz:2003xc, Creutz:2003xu, Creutz:2013xfa, Aoki:2014moa}. 
\hl{When $\eta'$ is included, however, the chiral Lagrangian has an ambiguity about its mass term; two conventional options are the Kobayashi-Maskawa-'t~Hooft determinant vertex, $(\rme^{-\im \theta}\operatorname{det}U+\mathrm{c.c.})$~\cite{Kobayashi:1970ji, Kobayashi:1971qz, Maskawa:1974vs, tHooft:1976rip}, and the log-det vertex, $(\im \ln \det(U)+\theta)^2$, supported by the large-$N$ analysis~\cite{Witten:1980sp, DiVecchia:1980yfw, Sakai:2004cn}. 
The global aspects of the phase diagram depend on the choice of the $\eta'$ mass term, which has led to some confusion. As a related problem, if we naively add the $\eta'$ meson to the chiral Lagrangian, such a model has struggled to reconcile with existing knowledge about the global structure, including the discrete anomalies of Refs.~\cite{Gaiotto:2017tne, Tanizaki:2017mtm, Tanizaki:2018wtg, Anber:2019nze} and the vacuum structure at the quenched limit ($m_u,m_d\to \infty$). 

This paper aims to provide a fresh perspective on these topics based on our semiclassical framework. 
Compared to the conventional chiral Lagrangian, our framework explicitly derives the global structure of $\eta'$ degrees of freedom from the microscopic calculation. 
We will see how our improvement of the $\eta'$ field settles the subtle issues on the phase structure in previous studies.
}

% The structure of this paper is organized as follows.
% In the subsequent subsection, we provide a summary of the main findings and their implications.
% In Section \ref{sec:adiabatic_continuity}, we revisit the semiclassical approach using the 't~Hooft flux $T^2$-compactification, focusing on the case of $SU(N)$ pure Yang-Mills theory as an example.
% Section \ref{sec:semiclassics_basics} details the setup for QCD and explains the derivation of the two-dimensional (2d) effective model after the small $T^2$ compactification.
% In Section \ref{sec:lesson}, we discuss the implications of the semiclassical 2D description on the chiral Lagrangian, particularly in relation to the $\eta'$ meson.
% Sections \ref{sec:NF-1} and \ref{sec:multiflavor} are dedicated to examining the vacuum structures in one-flavor and multi-flavor QCD, respectively, as inferred from the semiclassical method.
% Lastly, in Section \ref{sec:Dashen_phase}, we explore the Dashen phases in $N_f = 1+1$ QCD (and in $N_f = 1+1+1$ QCD with a constant strange quark mass) within the semiclassical framework.

\subsection{Summary and implications}

Let us summarize the setup, results, and implications.
We consider the $N_f$-flavor fundamental QCD, and take the $T^2$ compactification with the 't~Hooft and magnetic fluxes.
Because this semiclassical framework is designed to describe a confining vacuum, our study is limited to below the conformal window, $N_f < N_{\mathrm{CFT}}$.
%The setup will be described in detail in Section~\ref{sec:semiclassics_basics}, 
We obtain the following $2$d effective theory in the semiclassical regime (Sec.~\ref{sec:semiclassics_basics}):
\begin{itemize}
%     \item $N_f = 1$: The 2d effective description consists of a $2 \pi N$-periodic scalar $\eta'$, and its potential is,
% \begin{align}
%     V[\eta'] = - m \mu \cos(\eta') - 2 K \rme^{-S_I/N} \cos\left(\frac{\eta' - \theta}{N}\right), \tag{\ref{eq:one-flavor-potential}}
% \end{align}
% 
    \item[~] The 2d effective description consists of $2\pi N$-periodic scalar $\eta'$ and $(N_f - 1)$ $2\pi$-periodic scalars $\{ \varphi_1, \cdots ,\varphi_{N_f-1} \}$ with the potential,
\begin{align}
    V[\eta',\varphi_1, \cdots ,\varphi_{N_f-1}] &= - m \mu \cos(\eta' - (\varphi_1 + \cdots + \varphi_{N_f-1})) - m \mu \sum_j \cos(\varphi_j) \notag \\
    &~~~~~~~~~ - 2 K \rme^{-S_I/N} \cos\left(\frac{\eta' - \theta}{N}\right),  \tag{\ref{eq:potential_multiflavor}}
\end{align}
where $m$ is the quark mass, $\mu$ is a scale introduced by the bosonization scheme, and $K \rme^{-S_I/N}$ is the weight of the fractional instanton.
%When $N_f=1$, we just need to forget $\varphi_1,\ldots, \varphi_{N_f-1}$. 
If we adopt the non-Abelian bosonization, we obtain the 2d analog of the chiral Lagrangian (\ref{eq:NA_bosonized_effective}).
% \hl{The $N_f=1$ case simply consists of one $2\pi N$-periodic scalar $\eta'$ with the potential $V[\eta']$ (\ref{eq:one-flavor-potential}), which is just the above potential without $\{ \varphi_1, \cdots ,\varphi_{N_f-1} \}$.}

\end{itemize}
We here emphasize that the $\eta'$ field does not have the naive $2\pi$ periodicity but it has the extended $2\pi N$ periodicity. 
Under the assumption of the adiabatic continuity (see Fig.~\ref{fig:methodology}), the 2d effective theory at small $T^2$ can predict the vacuum structure of the original 4d QCD, and we propose the $4$d chiral Lagrangian with the $2\pi N$-periodic $\eta'$ field (Sec.~\ref{sec:lesson}).
We apply it to explain the QCD vacuum structures in the space of the fermion mass (Secs.~\ref{sec:NF-1} and~\ref{sec:multiflavor} for one-flavor and multi-flavor degenerate cases, respectively), and we also analyze the Dashen phases in $N_f = 1+1$ QCD and in $N_f = 1+1+1$ QCD with a constant strange quark mass (Sec.~\ref{sec:Dashen_phase}).

We highlight the following implications of our results:
\begin{itemize}
    \item The 2d analog of the chiral Lagrangian is explicitly derived from the microscopic computation based on QCD~\cite{Tanizaki:2022ngt}, not relying on the phenomenological Lagrangian. It can be used to study the $4$d QCD vacuum structure assuming the adiabatic continuity. 
    \item The derived 2d analog of the chiral Lagrangian naturally connects the chiral limit and the quenched limit, reproducing the widely accepted vacuum structure of QCD (Fig.~\ref{fig:multi_flavor_phase_diagram}). This, in turn, indirectly suggests the validity of the adiabatic continuity conjecture.
    \item 
    \hl{As mentioned above, the chiral Lagrangian suffers from an ambiguity when we try to introduce the $\eta'$ field.
    In our approach, this issue is resolved by clarifying that the $\eta'$ field has an extended periodicity of $2\pi N$, $\eta'\sim \eta'+2\pi N$, which is a consequence of integrating out the gluon's $\mathbb{Z}_N$ center contribution.}
    We can naturally fix the form of the $\eta'$ mass by the symmetry, locality, and dimensional counting, which proposes the improvement of $4$d chiral Lagrangian~\eqref{eq:4DchiralLagrangian_proposal}.  
    As an application, we revisit and refine the existence of the Dashen phase, deepening our understanding of these global aspects.
\end{itemize}

\section{Semiclassics for \texorpdfstring{$SU(N)$}{SU(N)} Yang-Mills theory: dilute gas of center vortices}
\label{sec:adiabatic_continuity}

This paper aims to explore the QCD vacuum structure and related physics using the semiclassical framework based on 't Hooft flux $T^2$ compactification, as proposed in Ref.~\cite{Tanizaki:2022ngt}. 
We here concisely summarize \hl{the methodology through the example of $SU(N)$ Yang-Mills theory}.

\hl{Based on the adiabatic continuity conjecture, we} investigate the physics of confining vacua in $\mathbb{R}^4$ from a weakly-coupled theory on $\mathbb{R}^2 \times T^2$ with a small torus $T^2$ \hl{(Figure~\ref{fig:methodology})}.
At small $T^2$, the theory becomes weakly-coupled and is described by a 2d low-energy effective theory. Here, quark confinement is explained by the dilute gas approximation of fractional instantons, namely center vortices.
\hl{The center-vortex picture is one of the plausible scenarios to explain quark confinement \cite{Greensite:2011zz}, 
and this semiclassical framework gives its explicit and calculable realization through the $T^2$ compactification.}
(For a similar approach with the dilute gas approximation of monopoles on $\mathbb{R}^3 \times S^1$, see, e.g., Refs.~\cite{Unsal:2007vu, Unsal:2007jx, Unsal:2008ch, Shifman:2008ja, Poppitz:2012sw, Poppitz:2012nz}.)

Let us review the semiclassical analysis of the $SU(N)$ Yang-Mills theory, and see how the dilute gas approximation of center vortices yields the confining vacuum \cite{Tanizaki:2022ngt}.
We consider the $SU(N)$ Yang-Mills theory on $\mathbb{R}^2 \times T^2$ at a small torus $T^2$, and we take the 't Hooft boundary condition for $T^2$.

Throughout this paper, we use the following coordinates: $x = (x_1,x_2,x_3,x_4) = (\Vec{x}, x_3, x_4) \in \mathbb{R}^2 \times T^2$, and we identify $(x_3,x_4) \sim (x_3+L,x_4) \sim (x_3,x_4+L)$ to form $T^2$.
The unit 't Hooft flux can be inserted by introducing the \hl{$SU(N)$-valued} transition functions $g_3(x_4)$ and $g_4(x_3)$ for $T^2$ satisfying
\begin{align}
    g_3(L)^\dagger g_4(0)^\dagger g_3(0) g_4(L) = \rme^{\frac{2 \pi \im}{N}} \hl{1_{N \times N}}.
\end{align}
We can take this boundary condition in the absence of charged matter of nontrivial $N$-ality.
In modern language, this boundary condition is equivalent to inserting the background field of the $\mathbb{Z}_N^{[1]}$ symmetry.

Up to a gauge transformation, one can regard the transition functions as the shift matrix $S$ and clock matrix $C$ of $SU(N)$
\begin{align}
    g_3(x_4) = S,~~~~g_4(x_3) = C, \label{eq:transition_functions_gauge}
\end{align}
with \hl{$C =\rme^{\im \alpha} \operatorname{diag}(1,\rme^{\frac{2 \pi \im}{N}}, \cdots,\rme^{\frac{2 \pi \im (N-1)}{N}})$, $(S)_{ij} =\rme^{\im \alpha} \delta_{i+1,j}$, and $\rme^{\im N \alpha}=(-1)^{N+1}$.}

We shall show that this setup provides the confining vacuum even at small $T^2$.
The important observations are listed as follows \cite{Yamazaki:2017ulc, Tanizaki:2022ngt, Cox:2019aji}.

\begin{itemize}
    \item \textbf{Center Symmetry}
    
By the $T^2$ compactification, the 1-form symmetry is decomposed as
\begin{align}
    \left( \mathbb{Z}_N^{[1]} \right)_{\mathrm{4d}} \rightarrow \left( \mathbb{Z}_N^{[1]}\right)_{\mathrm{2d}} \times \mathbb{Z}_N^{[0]} \times \mathbb{Z}_N^{[0]} , 
\end{align}
where $\mathbb{Z}_N^{[0]} \times \mathbb{Z}_N^{[0]}$ is the conventional center symmetry acting on the Polyakov loops of $T^2$.
\hl{From} the adiabatic continuity, the center symmetry $\mathbb{Z}_N^{[0]} \times \mathbb{Z}_N^{[0]}$ should not be spontaneously broken: the Polyakov loops along $x_3$ and $x_4$ directions should vanish $\braket{\operatorname{tr} P_3} =\braket{\operatorname{tr} P_4} = 0$.
With a periodic boundary condition, the center symmetry is spontaneously broken at a small compactification scale and the adiabatic continuity cannot be achieved\footnote{\hl{Without the 't~Hooft flux, the classical vacua are parametrized by the Polyakov loops $P_3, P_4$ with $P_3 P_4=P_4 P_3$, and the one-loop calculation produces the effective potential~\cite{Gross:1980br, Luscher:1982ma, Unsal:2010qh, Tanizaki:2022ngt},
\begin{align}
    V[P_3,P_4] = - \frac{2}{\pi^2 L^4} \sum_{(n_3,n_4) \neq (0,0)} \frac{1}{(n_3^2 + n_4^2)^2} \operatorname{tr} \left( P_3^{n_3}P_4^{n_4}\right).  \notag
\end{align}
This has $N^2$ minima located at $\frac{1}{N}\braket{\operatorname{tr} P_3} = \rme^{\frac{2 \pi m_3}{N}} ,~~\frac{1}{N}\braket{\operatorname{tr} P_4} = \rme^{\frac{2 \pi m_4}{N}}$, where $m_3, m_4 = 0,\cdots, N-1$, and $\mathbb{Z}_N^{[0]} \times \mathbb{Z}_N^{[0]}$ is spontaneously broken.  
}}. 
The 't Hooft boundary condition avoids this problem. Indeed, we can suppose that local fluctuations are negligible on a sufficiently small torus, and the transition functions give the Polyakov loops. In the gauge (\ref{eq:transition_functions_gauge}), their classical values are
\begin{align}
    P_3 = S,~~~ P_4 = C,
\end{align}
which satisfy $\braket{\operatorname{tr} P_3} =\braket{\operatorname{tr} P_4} = 0$.

    \item \textbf{Gapped gluons}

    With the choice of the transition functions (\ref{eq:transition_functions_gauge}), the boundary conditions for \hl{the gauge field} are given by
\begin{align}
    a_{\mu} (\Vec{x},x_3+L,x_4) = S a_{\mu}(\Vec{x},x_3,x_4) S^\dagger, ~~~~ a_{\mu}(\Vec{x},x_3,x_4+L) = C a_{\mu}(\Vec{x},x_3,x_4) C^\dagger,
\end{align}
which admits no constant mode.
Thus, the gluon field acquires a Kalzua-Klein \hl{mass of order $O(1/NL)$} without zero modes.
    \item \textbf{Center vortices/fractional instantons}
    
Let us suppose that we further compactify the $(x_1,x_2)$ directions into another torus with a nontrivial 't Hooft flux.
Then, the $\left( \mathbb{Z}_N^{[1]} \right)_{\mathrm{4d}}$ background $B$ representing the 't~Hooft twists has the self-intersection, say $\frac{N^2}{8\pi^2} \int B \wedge B  = 1$.
The topological charge is then,
\begin{align}
    Q_{\mathrm{top}} = \frac{1}{8\pi^2} \int_{\hl{T^2 \times T^2}} \operatorname{tr} \left( (\Tilde{f} - B) \wedge (\Tilde{f} - B)\right) \in - \frac{1}{N} + \mathbb{Z},
\end{align}
where $\Tilde{f}$ is the $U(N)$ field strength promoted from the original $SU(N)$ field strength $f$ satisfying $\operatorname{tr} \Tilde{f} = NB$.
Here, Bogomol'nyi–Prasad–Sommerfield (BPS) bound shows
\begin{align}
    S_{YM} = \frac{1}{g^2} \int \operatorname{tr}(f\wedge \star f) \geq \frac{8\pi^2}{Ng^2} = \frac{S_I}{N},
\end{align}
where $S_I=8\pi^2/g^2$ is the instanton action, and the equality holds if and only if the $f$ is (anti-)self-dual.
Some numerical studies \cite{Gonzalez-Arroyo:1998hjb, Montero:1999by, Montero:2000pb} support that the anti-self-dual solution indeed exists and this fractional instanton will survive as a local solution in the decompactification limit of $(x_1,x_2)$ directions.
\hl{
Based on these numerical observations, we assume that such a solution saturating the BPS bound with $Q_{\mathrm{top}} = \pm 1/N$ exists locally in our $\mathbb{R}^2 \times T^2$ setup}\footnote{\hl{Here, ``exists locally'' means that the solution exists if the boundary conditions at infinity in $\mathbb{R}^2$ are not considered.
Note that such a solution with fractional topological charge does not exist globally when the point at infinity of $\mathbb{R}^2$ is regular, i.e., when $\mathbb{R}^2$ can be compactified without any twist. Thus, in order to compute the partition function by the dilute gas approximation, we will consider ensembles of center vortices and anti-vortices, provided that the \textit{total} topological charge is an integer.
}}, which we call center vortex or fractional instanton\footnote{\hl{About the terminology, the ``fractional instanton'' would be obvious as it has the $\frac{1}{N}$ fractional topological charge compared with the minimal instanton on $S^4$. On the other hand, the ``center vortex'' may not be obvious as it should be defined by the following feature: When the vortex goes across the Wilson loop, its phase is rotated by the $\mathbb{Z}_N$ center element. In the $T^2$-compactified setup with the 't~Hooft flux, it turns out that the fractional instanton describes the center vortex~\cite{Gonzalez-Arroyo:1998hjb, Montero:1999by, Montero:2000pb}, so we identify them in this paper. }}. 
Similar fractional instantons in $\mathbb{R} \times T^3$ are also numerically studied in \cite{GarciaPerez:1989gt, GarciaPerez:1992fj, Itou:2018wkm}, which are also used in the different but deeply-related semiclassics~\cite{Yamazaki:2017ulc, Cox:2019aji}. 
For a recent review of the fractional instanton picture of confinement, see also \cite{Gonzalez-Arroyo:2023kqv}.
\end{itemize}

Based on these observations, the dilute gas approximation of center vortices provides a reasonable picture of the confining vacuum.
\hl{We consider ensembles of center vortices with $S = S_I/N,~Q_{\mathrm{top}} = 1/N$ and anti-vortices with $S = S_I/N,~Q_{\mathrm{top}} = -1/N$.}
The center-vortex and anti-center-vortex amplitudes can be expressed as
\begin{align}
    K \rme^{-S_I/N} \rme^{\im \theta / N},~~K \rme^{-S_I/N} \rme^{-\im \theta / N},
\end{align}
where $K$ is a dimensionful constant.
Although the center vortex can exist locally, the proper boundary condition of $\mathbb{R}^2$ forces the total topological charge $Q_{\mathrm{top}}$ to be an integer\footnote{More precisely, we here suppose that $\mathbb{R}^2$ is compactified without a 't Hooft flux. This is natural when we calculate the partition function. When the 't~Hooft flux is introduced also in the $\mathbb{R}^2$ direction, we can reproduce the anomaly for the $\theta$ periodicity~\cite{Tanizaki:2022ngt}. }.
The dilute gas approximation yields,
\begin{align}
    Z &= \sum_{n,\Bar{n} \geq 0} \frac{\delta_{n-\Bar{n}\in N \mathbb{Z}}}{n! \Bar{n}!} \left( K V \rme^{-S_I/N} \rme^{\im \theta / N}\right)^n \left(  K V \rme^{-S_I/N} \rme^{-\im \theta / N} \right)^{\Bar{n}} \notag \\
    &= \sum_{k \in \mathbb{Z}_N} \exp \left[2V K \rme^{-S_I/N} \cos \left( \frac{\theta - 2 \pi k}{N} \right) \right] \label{eq:parititon_fct_YM}
\end{align}
where $V$ is the volume in the $(x_1,x_2)$ directions.
From this partition function (\ref{eq:parititon_fct_YM}), we can deduce that there are $N$ vacua with the following energy density,
\begin{align}
    \mathcal{E}_k(\theta) = - 2 K \rme^{-S_I/N} \cos \left( \frac{\theta - 2 \pi k}{N} \right), ~~~k = 0 ,\cdots , N-1. \label{eq:vac_ene_YM}
\end{align}
At $\theta = 0$, the $k=0$ vacuum is the ground state.
As we increase $\theta$, the two-fold degenerate vacua appear at $\theta = \pi$, and the $k=1$ vacuum becomes the ground state for $\pi < \theta<3\pi$ via the level crossing.
This picture indeed explains the multi-branch structure of the Yang-Mills vacuum, which has been well-established in the large-$N$ limit~\cite{Witten:1980sp, Witten:1998uka, DiVecchia:1980yfw, DiVecchia:2017xpu}, in the dual superconductor picture~\cite{tHooft:1981bkw, Cardy:1981qy, Cardy:1981fd, Honda:2020txe, Hayashi:2022fkw}, and also in the semiclassics on $\mathbb{R}^3\times S^1$ with suitable deformations~\cite{Unsal:2007vu, Unsal:2007jx, Unsal:2008ch, Shifman:2008ja}.
One can also derive the area-law falloff of the Wilson loop from the dilute gas calculation \cite{Tanizaki:2022ngt}.

For a later purpose, let us conclude this section with a remark on the domain wall at $\theta = \pi$ (see also Sec.~5 of \cite{Hayashi:2023wwi}).
In the original 4d Yang-Mills theory, the domain wall should be a dynamical object equipped with the Chern-Simons theory~\cite{Gaiotto:2017yup}.
On the other hand, in the 2d effective description, the degenerate vacua are distinguished by the discrete label $k$, and the domain wall is a non-dynamical object.
Indeed, the domain wall connecting $k=0$ and $k=1$ vacua can be expressed by the external Wilson loop, which is a non-dynamical object.

To understand this difference before and after compactification, we have to note that the adiabatic continuity conjecture only claims the continuity of the vacuum and does not necessarily preserve excitations.
The dynamical domain wall becomes heavy as the torus is taken smaller.
In the small torus limit, the dynamical domain wall is integrated out, and no dynamical object connects the different vacua (\ref{eq:parititon_fct_YM}).
If we want to discuss dynamical domain walls in this framework, we need to consider the neglected Kalzua-Klein modes.

\section{QCD on \texorpdfstring{$\mathbb{R}^2 \times T^2$}{R2xT2} with 't Hooft and magnetic baryon fluxes}
\label{sec:semiclassics_basics}

\begin{figure}[t]
\centering
\includegraphics[scale=0.6]{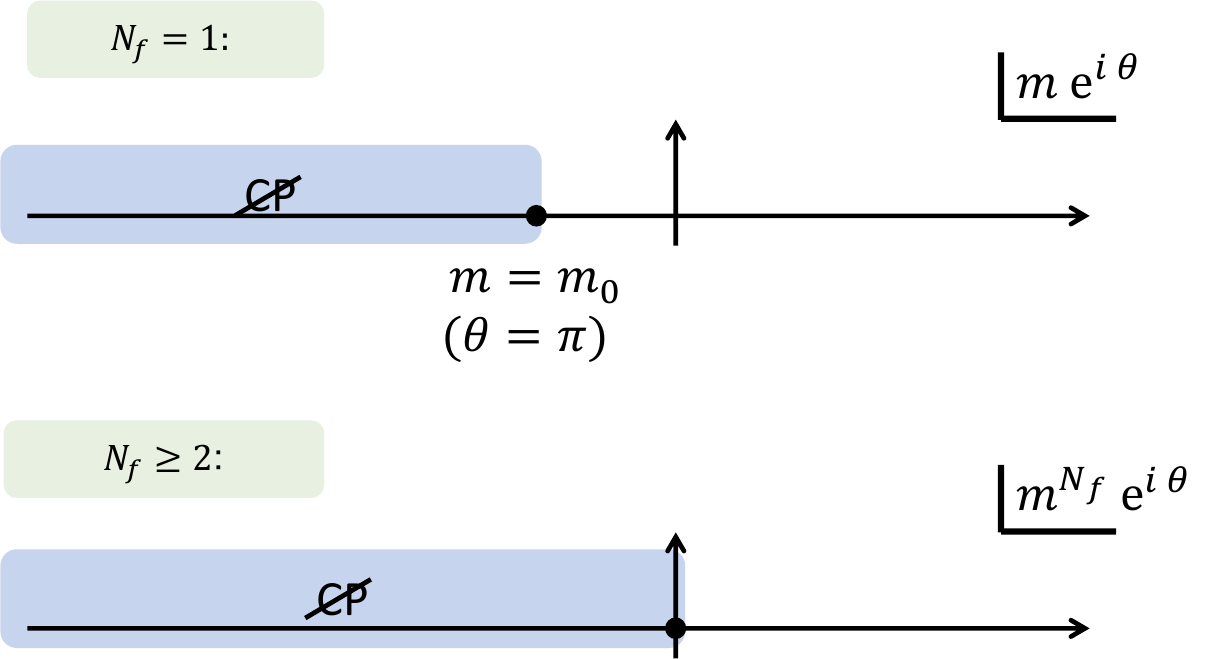}
\caption{Expected phase diagrams on complex quark mass, indicated by the large-$N$ argument \cite{Rosenzweig:1979ay, Nath:1979ik, Witten:1980sp, DiVecchia:1980yfw}:
(top panel)  for the one-flavor QCD, (bottom panel) for QCD with $2 \le N_f (< N_{\mathrm{CFT}})$ degenerate flavors.}
\label{fig:multi_flavor_phase_diagram}
\end{figure}

In this section, we focus on the flavor-symmetric QCD, consisting of the $SU(N)$ gauge field and $N_f$ fundamental quarks with mass $m$.
Here we take the degenerate quark mass to be positive, $m>0$, and the $CP$-violating phase is put to the $\theta$ angle. 
Since the semiclassical framework is constructed to describe a confining vacuum, we restrict ourselves below the conformal window $N_f < N_{\mathrm{CFT}}$.
The main purpose of this section is to explain the widely believed phase diagram in QCD with the vacuum angle $\theta$, depicted in Figure~\ref{fig:multi_flavor_phase_diagram} \cite{Rosenzweig:1979ay, Nath:1979ik, Witten:1980sp, DiVecchia:1980yfw} (see also \cite{Gaiotto:2017tne}), by the semiclassical approach of \cite{Tanizaki:2022ngt}.

\subsection{Setup}

We study $SU(N)$ gauge theory with $N_f$ fundamental quarks of degenerate mass $m$.
Based on the adiabatic continuity conjecture, we compactify $\mathbb{R}^4$ into $\mathbb{R}^2 \times T^2$ with the 't Hooft and baryon fluxes.
This compactification was employed in Section 4.2 of \cite{Tanizaki:2022ngt}, and we will expand upon this semiclassical analysis, focusing specifically on the dependence of the $\theta$ angle. 
For completeness, let us briefly review the setup.

We compactify the spacetime manifold into $\mathbb{R}^2 \times T^2$ with a minimal 't Hooft flux for $SU(N)$.
As above, we can introduce the minimal 't Hooft flux by taking the transition functions $g_3(x_4)$ and $g_4(x_3)$ for $T^2$ satisfying
\begin{align}
    g_3(L)^\dagger g_4(0)^\dagger g_3(0) g_4(L) = \rme^{\frac{2 \pi \im}{N}} 1_{N\times N}.
\end{align}
Unlike the pure Yang-Mills case, in the presence of a fundamental-representation fermion, the nontrivial 't Hooft flux gives an inconsistency. 
\hl{Indeed, for a fundamental quark $\psi(x_3,x_4)$, there are two ways to connect $\psi(0,0)$ to $\psi(L,L)$ using transition functions: $\psi(0,0) = g_3 (0) g_4 (L) \psi(L,L) = g_4 (0) g_3 (L) \psi(L,L)$, which implies $g_3(L)^\dagger g_4(0)^\dagger g_3(0) g_4(L) = 1_{N\times N}$. We cannot introduce the 't Hooft flux alone in this case.}
To resolve this problem, we need to remember that the symmetry group has the nontrivial quotient with the gauge redundancy, 
\begin{align}
    G_{\mathrm{gauge}+\mathrm{global}}=\frac{SU(N)_{\gauge} \times SU(N_f)_V \times U(1)_V}{\mathbb{Z}_N \times \mathbb{Z}_{N_f}}. 
    \label{eq:symmetrygroup}
\end{align}
To see how it works, let us introduce the unit flux of the baryon-number symmetry $U(1)_B$ in the torus $T^2$, which is the $1/N$ fractional flux in terms of the quark-number symmetry $U(1)_V$ as $U(1)_B=U(1)_V/\mathbb{Z}_N$:
\begin{align}
    \int_{T^2} \diff A_B = 2\pi.
\end{align}
For example, we can take 
\begin{align}
    A_B = \frac{2 \pi}{L^2} x_3 \diff x_4. \label{eq:bayon_gauge_bg}
\end{align}
In this gauge field configuration, the $U(1)_B$ magnetic flux can be understood from the non-trivial $U(1)_B$ transition function: $\rme^{\im \alpha(x_4)} = \rme^{\im \frac{2 \pi x_4}{L}}$ arising from
\begin{align}
    A_B(x_3+L,x_4) = A_B(x_3,x_4) + \frac{2 \pi}{L} \diff x_4.
\end{align}
In the setup with both $SU(N)_{\rm{gauge}}$ 't Hooft flux and $U(1)_B$ magnetic flux, the boundary conditions for the fundamental quarks $\psi(x)$ are
\begin{align}
    \psi(\Vec{x}, x_3+L, x_4) &= \rme^{-\im \alpha(x_4)/N} g_3(x_4)^\dagger \psi(\Vec{x}, x_3, x_4)   \notag \\
    \psi(\Vec{x}, x_3, x_4+L) &=  g_4(x_3)^\dagger \psi(\Vec{x}, x_3, x_4),
\end{align}
which are now consistent with the single-valuedness.
This pair of the 't Hooft and $U(1)_B$ fluxes can be interpreted by the 2-form background associated with the $\mathbb{Z}_N$ quotient of the symmetry group~\eqref{eq:symmetrygroup}.

\subsection{2d effective description}
\label{sec:2d_EFT}

Now, we consider the 2d low-energy description of the QCD on $\mathbb{R}^2 \times T^2$ with a small size of torus $L \ll \Lambda_{QCD}^{-1}$.
Under the assumption of the adiabatic continuity, this 2d effective theory can predict the original QCD phase structure on $\mathbb{R}^4$.

The procedure to derive the 2d effective theory consists of the following two steps:
\begin{enumerate}
     \item Identify the ``low-energy modes,'' which are perturbatively massless after the small $T^2$ compactification.
    \item Incorporate center vortices by the dilute gas approximation
\end{enumerate}

\subsubsection{Low-energy modes}
In what follows, we choose the gauge so that the transition functions are the shift matrix $S$ and clock matrix $C$ of $SU(N)$
\begin{align}
    g_3(x_4) = S,~~~~g_4(x_3) = C. 
\end{align}
To derive the 2d effective theory, let us determine low-energy modes on small $T^2$.
For \hl{the gauge field}, the boundary conditions are
\begin{align}
    a_{\mu} (\Vec{x},x_3+L,x_4) = S a_{\mu}(\Vec{x},x_3,x_4) S^\dagger, ~~~~ a_{\mu}(\Vec{x},x_3,x_4+L) = C a_{\mu}(\Vec{x},x_3,x_4) C^\dagger, 
    \label{eq:bdy_cond_gauge}
\end{align}
which admit no constant mode in $\mathfrak{su}(N)$, and all perturbative excitations become massive with the gap of order $O(1/NL)$\footnote{We do not expect the adiabatic continuity for phases with gapless gluonic degrees of freedom, so we restrict ourselves to $N_f < N_{\mathrm{CFT}}$.}.
This perturbative mass gap of gluons can be understood as Higgsing by the Polyakov loops \cite{Tanizaki:2022ngt}:
\begin{align}
    SU(N) \xrightarrow{\mathrm{Higgsing}} \mathbb{Z}_N.
\end{align}

For quarks, the boundary conditions are
\begin{align}
    \psi(\Vec{x}, x_3+L, x_4) &= \rme^{-\im \frac{2 \pi x_4}{NL}} S^\dagger \psi(\Vec{x}, x_3, x_4)   \notag \\
    \psi(\Vec{x}, x_3, x_4+L) &=  C^\dagger \psi(\Vec{x}, x_3, x_4). \label{eq:bdy_cond_quarks}
\end{align}
The low-energy mode can be found by solving the zero-mode equation:
\begin{align}
    \left[ \gamma_3 \partial_3 + \gamma_4 \left( \partial_4 + \im \frac{1}{N} A_{B,4} \right) \right] \psi(x_3, x_4) = 0.
\end{align}
whose solutions give perturbatively massless modes.
Immediately, we can see the normalizable zero modes should have the positive eigenvalue of $\im \gamma_3 \gamma_4$: $\im \gamma_3 \gamma_4 \psi = \psi$, which reduces a 4d spinor to a 2d spinor.
After some computations shown in \cite{Tanizaki:2022ngt}, we obtain one zero-mode per flavor with the boundary condition (\ref{eq:bdy_cond_quarks}).
\hl{One can also determine the number of zero-modes by applying the index theorem on $T^2$.}
We thus have $N_f$ 2d Dirac fermions as 2d low-energy modes.
For later convenience, we bosonize these fermions in the 2d effective theory.

For $N_f = 1$, we use the Abelian bosonization.
The 2d free fermion is translated to the Sine-Gordon model described by a compact scalar $\varphi \sim \varphi + 2\pi$:
\begin{align}
    S[\varphi] =  \int \diff^2x \left[ \frac{1}{8\pi} (\partial \varphi)^2 - m \mu \cos \varphi \right],
\end{align}
with a dimensionful constant $\mu$, depending on the renormalization scheme.

We have two options of bosonization when $N_f \geq 2$, the non-Abelian bosonization and the Abelian bosonization for each component.
Although the non-Abelian bosonization, which gives $U(N_f)_1$ Wess-Zumino-Witten theory, would be better to keep the flavor symmetry manifest, here we adopt the Abelian bosonization to simplify some calculations.
With this scheme, we have $N_f$ compact bosons $\varphi_j \sim \varphi_j + 2\pi$ ($j = 1 ,\cdots, N_f$) and the action is,
\begin{align}
    S[\varphi_1,\cdots, \varphi_{N_f}] = \sum_{j=1}^{N_f} \int \diff^2x \left[ \frac{1}{8\pi} (\partial \varphi_j)^2 - m \mu \cos \varphi_j \right].
\end{align}

\subsubsection{Center vortices and residual \texorpdfstring{$\mathbb{Z}_N$}{ZN} gauge field}
\label{sec:QCD_center_vortex}

In order to incorporate center-vortex contribution, we have to determine the center-vortex vertex in the presence of quarks.

It is easy to determine the center-vortex vertex from the $U(1)_{\mathrm{axial}}$ spurious symmetry: for $N_f = 1$
\begin{align}
    \theta \rightarrow \theta + 2 \alpha,~~~\varphi \rightarrow \varphi + 2 \alpha,
\end{align}
and for $N_f \geq 2$,
\begin{align}
    \theta \rightarrow \theta + 2 \alpha,~~~\varphi_j \rightarrow \varphi_j + 2 \alpha/N_f ~~(j=1,\cdots,N_f).
\end{align}
Then, the center-vortex amplitude can be uniquely determined by the spurious symmetry: For $N_f = 1$,
\begin{align}
    \mathcal{V}(\Vec{x}) = ``[K \rme^{-S_I/N} \rme^{\im \theta /N}  \rme^{- \im \varphi /N}]," \label{eq:center-vortex-one-flavor}
\end{align}
where $S_I$ is the instanton action and $K$ is some appropriate (dimensionful) constant. \hl{We put the double quotations to indicate that careful consideration is required to define the fractional vertex $\rme^{- \im \varphi /N}$, which is discussed shortly thereafter.}
For $N_f \geq 2$, the center-vortex vertex becomes\footnote{With the non-Abelian bosonization, the center-vortex vertex should be 
\begin{align}
    \mathcal{V}(\Vec{x}) = ``[K \rme^{-S_I/N} \rme^{i \theta /N}  (\operatorname{det}U)^{-1/N}],"  
\end{align}
where $U \in U(N_f)$ is the non-Abelian bosonized variable. This is nothing but the $1/N$ fractionalization of the Kobayashi-Maskawa-'t~Hooft instanton vertex~\cite{Kobayashi:1970ji, Kobayashi:1971qz, Maskawa:1974vs, tHooft:1976rip}. }
\begin{align}
    \mathcal{V}(\Vec{x}) = ``[K \rme^{-S_I/N} \rme^{\im \theta /N} \rme^{- \im (\varphi_1 + \cdots + \varphi_{N_f}) /N}],"
\end{align}
where we also use the flavor permutation symmetry, a subgroup of $SU(N_f)$.
At first, it appears that they are non-genuine operators as they violate the periodicity of compact bosons, and topological lines attached to them would be required. However, we can resolve this problem by remembering that there is a $\mathbb{Z}_N$ gauge field.
The key observations are the following two properties:
\begin{itemize}
    \item There is $\mathbb{Z}_N$ gauge field after the Higgsing by the Polyakov loops.
    Since the original fermion is in the fundamental representation, the residual $\mathbb{Z}_N$ gauge symmetry is the vector-like symmetry $\mathbb{Z}_N \subset U(1)_V$.
    \hl{In the bosonization dictionary, the fermion vector-like symmetry is mapped into the magnetic (winding) symmetry of the bosonized variable.
    The fermion number is translated as the number of kinks, which is the winding number.} 
    Thus, after the bosonization, the coupling should be\footnote{Note that this magnetic coupling is consistent with the center-vortex vertex (\ref{eq:center-vortex-one-flavor}). Intuitively, the center vortex introduces a fractional magnetic flux $``\diff a = \frac{2\pi}{N} \delta (\vec{x})"$, which leads to the expression (\ref{eq:center-vortex-one-flavor}) from the magnetic coupling (\ref{eq:Z_N_residual_gauge}).}
    \begin{align}
        \frac{\im}{2\pi} \int a_{\mathbb{Z}_N} \wedge \diff \varphi , \label{eq:Z_N_residual_gauge}
    \end{align}
    or
        \begin{align}
        \frac{\im}{2\pi} \int a_{\mathbb{Z}_N} \wedge \diff (\varphi_1 + \cdots + \varphi_{N_f}) . 
    \end{align}
    
    \item There is a constraint that the \hl{\textit{total}} topological charge must be an integer. 
    This constraint leads to the sum over the vacuum label $k \in \mathbb{Z}_N$, as seen in (\ref{eq:parititon_fct_YM}).
\end{itemize}
These two properties leads to the extension of $\varphi$ (or $\varphi_1 + \cdots + \varphi_{N_f}$) when integrating out gluon's center contributions, and we denote the extended $2\pi N$-periodic fields as $\eta'$:\footnote{The extension happens also when we perform the non-Abelian bosonization. In that case, the bosonized variable is a $U(N_f)$-valued field, which can be expressed as $(\varphi, \Tilde{U}) \in \mathbb{R} \times SU(N_f)$ with the identification $(\varphi, \Tilde{U}) \sim (\varphi + 2 \pi, \rme^{ -\frac{2 \pi \im}{N_f}} \Tilde{U})$. The phase of $\operatorname{det} U$ is given by $ \varphi$ as $(\varphi, \Tilde{U})\mapsto U=\rme^{\im \varphi/N_f}\Tilde{U} \in U(N_f)$.
After the $\mathbb{Z}_N$ gauging, the phase of $\operatorname{det} U$ becomes $2 \pi N$-periodic $\eta'$. Then, the identification is modified as $(\eta', \Tilde{U}) \sim (\eta' + 2 \pi N, \rme^{ -\frac{2 \pi N\im}{N_f}} \Tilde{U})$, and the chiral effective field is no longer $U(N_f)$-valued. },\footnote{This extension says that the $2\pi N$-winding vortex is minimal as the genuinely codim-$2$ object, but we would like to note that the $2\pi$-winding vortex also exists as a boundary of one-higher-dimensional object. Indeed, the $2\pi$-winding $\eta'$ vortex violates the $\mathbb{Z}_N$ gauge invariance due to the magnetic coupling \eqref{eq:Z_N_residual_gauge}, so it has to be an endpoint of the open $\mathbb{Z}_N$ Wilson loop to be gauge invariant. This is the $2$d counterpart of the ``pancake'' baryon proposed in Ref.~\cite{Komargodski:2018odf}, where the $2\pi$-winding $\eta'$ vortex lives at the edge of the Chern-Simons-enriched domain wall. More intuitively, we may say that $\eta'$ extends its periodicity by ``eating'' the $\mathbb{Z}_N$ label for the Yang-Mills confining vacua, so it must jump by $2\pi$ when the original vacuum label jumps at some dislocation. This would clarify why the ``pancake'' is needed in the proposal of \cite{Komargodski:2018odf}.
\label{ftnt:2piNextension}}
\begin{equation}
    \varphi\sim \varphi+2\pi \xRightarrow{\int \Diff a_{\mathbb{Z}_N}} \eta' \sim \eta'+2\pi N. 
\end{equation}

This claim can be explained as follows:
As mentioned above, the operator $\rme^{- \im \varphi /N}$ seems non-genuine, attached by a $\mathbb{Z}_N$ line.
Here, the residual $\mathbb{Z}_N$ gauge magnetically couples to $\varphi$, and the gauging of the magnetic $\mathbb{Z}_N$ transforms the non-genuine charge-$1/N$ electric object into a genuine operator, thereby validating its definition. 
More specifically, a $2\pi$ periodic compact scalar defined on $M_2$ is represented as follows:
\begin{align}
\diff \varphi = \diff \Tilde{\varphi} + c,
\end{align}
where
\begin{itemize}
    \item ``local fluctuations'': a globally-defined $\Tilde{\varphi} \in \mathbb{R}$,
    \item ``topological sectors'': $c/2\pi \in H^1 (M_2;\mathbb{Z})$.
\end{itemize}
Thus, the gauging of magnetic $\mathbb{Z}_N$ imposes a restriction, leading to $c=Nc'$ with some $c'/2\pi  \in H^1 (M_2;\mathbb{Z})$, and then 
\begin{equation}
    \diff \eta':=\diff \varphi=\diff \Tilde{\varphi}+Nc'
\end{equation}
defines the $2\pi N$-periodic compact scalar $\eta'$. Additionally, when we lift $\varphi\sim \varphi+2\pi$ to $\eta'\sim \eta'+2\pi N$, there is an ambiguity to add a discrete constant $2 \pi k ~(k \in \mathbb{Z}_N)$. Namely, the globally-defined $\Tilde{\varphi} \in \mathbb{R}$ has a $1$-to-$N$ correspondence between $\mathbb{R}/2\pi \mathbb{Z}$ and $\mathbb{R}/2\pi N \mathbb{Z}$. 
This $1$-to-$N$ correspondence absorbs the vacuum label $k \in \mathbb{Z}_N$ in the dilute-gas computation of center vortices.
Hence, the periodicity is extended due to the gauging of magnetic $\mathbb{Z}_N$ and the absorption of the Yang-Mills vacuum label.
We also give its clear demonstration in Appendix~\ref{sec:extendedperiodicity} using the bosonized Schwinger model. 

\hl{To be concrete, let us explicitly see how the $2$d effective theory is derived for $N_f=1$ case. As in (\ref{eq:parititon_fct_YM}), the dilute gas approximation with residual $\mathbb{Z}_N$ gauging yields,
\begin{align}
    Z &= \int \Diff a_{\mathbb{Z}_N} \int \mathcal{D}\varphi  \sum_{k \in \mathbb{Z}_N}  \rme^{\frac{\im}{2\pi} \int a_{\mathbb{Z}_N} \wedge \diff \varphi } \notag \\
    &~~~~~ \times \exp \left[ - S[\varphi]   + \int d^2 \Vec{x} \left\{ 2 K \rme^{-S_I/N} \cos \left( \frac{\varphi(\Vec{x}) - 2 \pi k}{N} \right) \right\} \right]
\end{align}
As explained above, the integration over the residual gauge $a_{\mathbb{Z}_N}$ constrains the winding number of $\varphi$: $\int \diff \varphi \in 2 \pi N \mathbb{Z}$, which allows us to lift $\varphi$ to a $2 \pi N$-periodic variable.
Here, there is $\mathbb{Z}_N$ ambiguity to lift $\varphi$: for given $\varphi$, we have $N$ corresponding $\mathbb{R}/2 \pi N \mathbb{Z}$ variable $\varphi \rightarrow \eta' =  \varphi - 2 \pi k~~(k \in \mathbb{Z}_N)$.
Based on this 1-to-$N$ correspondence, we notice the identity:
\begin{align}
    \int_{\eta' \sim \eta' + 2 \pi N} \mathcal{D}\eta' ~  (\cdots) = \int_{\left[ \frac{\diff \varphi}{2 \pi} \right] \in N H^1 (M_2;\mathbb{Z})} \mathcal{D}\varphi \sum_{k \in \mathbb{Z}_N} \left.(\cdots) \right|_{\eta' = \varphi - 2 \pi k},
\end{align}
where $\left[ \frac{\diff \varphi}{2 \pi} \right] \in N H^1 (M_2;\mathbb{Z})$ means the constraint that  the winding number of $\varphi$ is a multiple of $N$.
This identity enables us to rewrite the partition function neatly:
\begin{align}
    \int \Diff a_{\mathbb{Z}_N} \int \mathcal{D}\varphi  \sum_{k \in \mathbb{Z}_N}  \rme^{\frac{\im}{2\pi} \int a_{\mathbb{Z}_N} \wedge \diff \varphi } (\cdots) \Rightarrow \int_{\eta' \sim \eta' + 2 \pi N} \mathcal{D}\eta' 
    \left.(\cdots) \right|_{\varphi - 2 \pi k \mapsto \eta'}.
\end{align}
}

We now find the following 2d effective theory for QCD on $\mathbb{R}^2 \times T^2$ with 't~Hooft flux:
\begin{itemize}
    \item $N_f = 1$: we have a compact scalar with the periodicity $\eta' \sim \eta' + 2 \pi N$. The potential for this scalar is
\begin{align}
    V[\eta'] = - m \mu \cos(\eta') - 2 K \rme^{-S_I/N} \cos\left(\frac{\eta' - \theta}{N}\right).  \label{eq:one-flavor-potential}
\end{align}
    \item $N_f \geq 2$: We have one extended $2\pi N$-periodic scalar $\eta'$ and $2\pi$-periodic scalars $\{ \varphi_1, \cdots ,\varphi_{N_f-1} \}$. The potential is
\begin{align}
    V[\eta',\varphi_1, \cdots ,\varphi_{N_f-1}] &= - m \mu \cos(\eta' - (\varphi_1 + \cdots + \varphi_{N_f-1})) - m \mu \sum_j \cos(\varphi_j) \notag \\
    &~~~~~~~~~ - 2 K \rme^{-S_I/N} \cos\left(\frac{\eta' - \theta}{N}\right). 
    \label{eq:potential_multiflavor}
\end{align}
\end{itemize}

\section{Lesson about the chiral Lagrangian with \texorpdfstring{$\eta'$}{eta prime} in \texorpdfstring{$4$}{4}d}
\label{sec:lesson}

Before proceeding to the semiclassical analyses, let us comment on the significance of the extended periodicity of the compact boson, and we discuss its implications on the chiral effective Lagrangian with the $\eta'$ meson in $4$ dimension.

The non-Abelian-bosonized 2d effective theory is strikingly similar to the chiral Lagrangian as observed in Ref.~\cite{Tanizaki:2022ngt} (despite the fact that the chiral symmetry cannot be spontaneously broken in 2d due to Coleman-Mermin-Wagner theorem\footnote{
This subtlety reflects that the theory on $\mathbb{R}^4$ is not always equivalent to the theory on $\mathbb{R}^2 \times T^2$ at large $T^2$ in the naive way when the system is gapless.
We have to take into account how the IR description is modified when we compactify $\mathbb{R}^4$ to $\mathbb{R}^2 \times T^2$ with a large-but-finite torus size.
In our case, as we well see in Section \ref{sec:NA_bosonization_chiral_lagrangian}, the chiral Lagrangian on $\mathbb{R}^2 \times T^2$ gives the $SU(N_f)_1$ WZW theory, so the adiabatic continuity between the theories at large and small $T^2$ appears to work well. }), 
and it is consistent with the $T^2$ flux compactification of the chiral Lagrangian (Sec.~\ref{sec:NA_bosonization_chiral_lagrangian})\footnote{
We note that the consistency holds both for $N\ge 3$ and $N=2$. Even for the two-color case ($N=2$), the Pauli-G\"ursey extended $SU(2N_f)$ symmetry reduces to the usual chiral symmetry $SU(N_f)_L\times SU(N_f)_R\times U(1)_V$ as we introduce the $U(1)_B$ flux $A_B$ along the $T^2$ direction.
% In the two-color QCD ($N=2$), the flavor symmetry is extended to the Pauli-G\"ursey $SU(2 N_f)$ symmetry. The symmetry-breaking pattern is $SU(2 N_f) \rightarrow Sp(2 N_f)$, and the chiral Lagrangian includes massless baryons as well as mesons.
% The semiclassical description, which corresponds to the usual chiral Lagrangian of mesons (\ref{eq:NA_bosonized_effective}), seems to lack these baryons.
% However, as our compactification scheme gives order-$O(1/L)$ mass to the baryons, the semiclassical effective theory is consistent with the $SU(2 N_f)/Sp(2 N_f)$ chiral Lagrangian on $\mathbb{R}^2 \times T^2$ in the presence of the 't Hooft and $U(1)_B$ magnetic fluxes.
% Indeed, in our setup, a diquark baryon field $d$ couples to the $U(1)_B$ gauge background $A_B = \frac{2 \pi}{L^2} x_3 \diff x_4$ (\ref{eq:bayon_gauge_bg}) with the boundary conditions $d(x_3 + L, x_4) = \rme^{ - \frac{2 \pi \im x_4}{L}} d(x_3, x_4),~ d(x_3, x_4+L) = d(x_3, x_4)$. There is no zeromode $\partial_3 d(x_3, x_4) = 0,~ \left(\partial_4 - \im \frac{2 \pi x_3}{L^2} \right) d(x_3, x_4) = 0$ satisfying the boundary conditions, so the baryon field acquires mass of order $O(1/L)$ in the $2$d description.
}.
With this identification, the extended $2\pi N$-periodic compact boson corresponds to the $\eta'$ meson.
Then, we see that the extension of periodicity is important to couple $\eta'$ to the $U(1)_B$ background also in $4$ dimensions (Sec.~\ref{sec:extended_periodicity_anomaly}), 
and we propose the $4$d chiral Lagrangian with extended $\eta'$ that can reproduce the discrete baryon-color-flavor anomaly of Refs.~\cite{Gaiotto:2017tne, Tanizaki:2017mtm, Tanizaki:2018wtg, Anber:2019nze}.
%The pivotal point is that the $\mathbb{Z}_N$ part of the gauge field should not be discarded even in the low-energy limit to correctly capture the global structure. 
%The extension of the $\eta'$ periodicity is a consequence of the $\mathbb{Z}_N$ gauge in the 2d effective theory, which plays an essential role in capturing the global structure.

\subsection{Non-abelian bosonization and chiral Lagrangian}
\label{sec:NA_bosonization_chiral_lagrangian}

In the non-Abelian bosonization scheme, the resulting 2d effective theory (on $M_2$) is the level-$1$ $U(N_f)$ Wess-Zumino-Witten (WZW) model with the mass and center-vortex deformations,
\begin{align}
    S[U] &= \frac{1}{8 \pi} \int_{M_2} \diff^2 x \left( \operatorname{tr}(\partial_\mu U^\dagger \partial_\mu U) + V(U) \right) + \frac{1}{12 \pi} \int_{M_3} \operatorname{tr}\left( (U^{-1} \diff U)^3 \right)\label{eq:NA_bosonized_effective}, \notag \\
    V(U) &=  - m \mu \left(\operatorname{tr} U +\mathrm{c.c.}\right) - K \rme^{-S_I/N} \left( \rme^{\im \theta /N}  (\operatorname{det}U)^{-1/N} +\mathrm{c.c.} \right), 
\end{align}
where the phase of $\operatorname{det} U$ is extended to have $2\pi N$ periodicity as above, and we set $\partial M_3 = M_2$.

As noted in \cite{Tanizaki:2022ngt}, this effective model is quite similar to the chiral Lagrangian with the $\eta'$ field, the would-be Nambu-Goldstone boson for anomalously-broken $U(1)_{\mathrm{axial}}$.
Indeed, the $SU(N_f)$ chiral field couples to the baryon background through 
\begin{align}
    \int_{M_5} \diff A_B \wedge \left( \frac{1}{24 \pi^2} \operatorname{tr} (U^{-1} \diff U)^3 \right), \label{eq:Skyrmion_5d}
\end{align}
in the 5d WZW-type notation. The $T^2$ compactification with a baryon magnetic flux reduces (\ref{eq:Skyrmion_5d}) to the 3d WZW term of (\ref{eq:NA_bosonized_effective}).
Therefore, the non-Abelian bosonized 2d effective theory is consistent with the standard chiral Lagrangian, up to $\eta'$, which is gapped out by center-vortex vertices.

\subsection{Extended periodicity, \texorpdfstring{$U(1)_B$}{U(1)B} background, and discrete anomaly}
\label{sec:extended_periodicity_anomaly}

To incorporate the pseudo-Nambu-Goldstone $\eta'$ field into the chiral Lagrangian, we often extend the $SU(N_f)$-valued field to $U(N_f)$-valued field with some ``mass term'' for the $\eta'$ mode.
In our formalism, the periodicity of this $\eta'$ mode is extended.
In this section, we shall see how the extension of periodicity of $\eta'$ is essential to correctly preserve the global structures, such as the discrete anomaly. 

In the first place, we determine the coupling of the $U(1)_B$ background to the chiral field with $\eta'$.
To this end, we first see the coupling between the $U(1)_V$ background $A_{U(1)}~(= A_B/N)$ and a $U(N_f)$ chiral field $U$. 
We can determine this coupling by observing the WZW term, which gives, \hl{in the 5d notation,
\begin{align}
\Gamma_{WZW} &\supset \frac{\im N }{24 \pi^2} \int_{M_5} \left\{ \diff A_{U(1)} \wedge \operatorname{tr} \left( U^{-1} \diff U \right)^3 + 3 ~ \diff A_{U(1)} \wedge  \diff A_{U(1)} \wedge \operatorname{tr} \left( U^{-1} \diff U \right)  \right\} \notag \\
&= \frac{\im }{24 \pi^2} \int_{M_5} \left\{ \diff A_B \wedge \operatorname{tr} \left( U^{-1} \diff U \right)^3 + \frac{3}{N} ~ \diff A_B \wedge  \diff A_B \wedge \diff ( \ln \operatorname{det} U)  \right\},
\end{align}
}where the trace is taken over the flavor indices.
This can also be obtained by superficially employing the Goldstone-Wilzeck baryon current (with the charge matrix $Q = \frac{1}{N} \textbf{1}_{N_f \times N_f}$) \cite{Witten:1983tw, Goldstone:1981kk}.
Here, the first term is the well-known Skyrmion current, which only couples to the $SU(N_f)$ part.
The second term describes the coupling of $\eta' ( = \ln \operatorname{det} U)$ to the baryon background.
The coupling to the baryon instanton number is natural because $\eta'$ is the pseudo-Nambu-Goldstone boson of anomalously-broken $U(1)_{\mathrm{axial}}$, which has an extra anomaly in the presence of the $U(1)_V$ background\footnote{This $1/N$ fractional coupling describes the fractional quantum Hall effect on the domain walls \cite{Komargodski:2018odf} (see also footnote~\ref{ftnt:2piNextension}). Moreover, upon the $T^2$ compactification with the nontrivial baryon flux, this coupling  reproduces \eqref{eq:Z_N_residual_gauge} of the 2d effective theory, 
\begin{align*}
\frac{\im}{8\pi^2 N}\int_{\mathbb{R}^2\times T^2} \eta' \diff A_B\wedge \diff A_B\Rightarrow \frac{\im}{2 \pi N} \int_{\mathbb{R}^2} \eta' \diff A_B= \frac{\im}{2 \pi} \int_{\mathbb{R}^2} \varphi\, \diff A_{U(1)}.
\end{align*}}. 
However, this coupling to $\eta'$,
\begin{align}
\frac{\im}{8 \pi^2 N} \int \eta' \diff A_B \wedge \diff A_B,
\label{eq:etaU1Bcoupling}
\end{align}
is naively ill-defined for $2 \pi$-periodic $\eta' $.
We need to extend the periodicity to make the $U(1)_B$ coupling well-defined.

We note that the baryon number symmetry $U(1)_B = U(1)_V/\mathbb{Z}_N$ has the $\mathbb{Z}_N$ quotient compared with the naive quark-number symmetry $U(1)_V$, and this originates from the quotient structure between the gauge group and the global symmetry group, $\frac{SU(N)_{\mathrm{gauge}} \times U(1)_V}{\mathbb{Z}_N}$. Thus, naively ignoring $SU(N)_{\mathrm{gauge}}$ makes the coupling with $U(1)_B = U(1)_V/\mathbb{Z}_N$ ill-defined. In fact, in the 2d effective theory, $(\mathbb{Z}_N)_{\mathrm{gauge}} \subset SU(N)_{\mathrm{gauge}}$ induces a periodicity extension, making the coupling with $U(1)_B = U(1)_V/\mathbb{Z}_N$ well-defined.
Hence, it is necessary to remember at least $(\mathbb{Z}_N)_{\mathrm{gauge}}$ to correctly couple to the $U(1)_B$ background.
In the 2d effective theory, this can be achieved by the periodicity extension\footnote{In this context, the role of vector mesons is sometimes emphasized in some literature \cite{Kitano:2020evx, Karasik:2020zyo}.
Certainly, the vector mesons can rectify the coupling with the background gauge fields because the vector mesons represent gluons screened by quarks.
Nevertheless, the punchline here is that we should not neglect $(\mathbb{Z}_N)_{\mathrm{gauge}}$, and its primary effect can be taken into account by extending the periodicity of $\eta'$ rather than including those heavy local excitations.}.

With the periodicity extension, $\eta' \sim \eta' + 2\pi N$, we shall see that not only the 2d effective theory but also the 4d chiral Lagrangian can reproduce the baryon-color-flavor anomaly, which exists at $\theta = \pi$ for $\operatorname{gcd}(N,N_f) \neq 1$.
We can also propose the $4$d chiral Lagrangian with such an extended $\eta'$ field as
\begin{align}
    S_{\mathrm{chiral}}[U]  \sim \int f_\pi^2 |\diff U|^2 - m\Lambda^3 (\operatorname{tr} U + \mathrm{c.c.})-\frac{N^2\chi_{\mathrm{YM}}}{2} \left( (\rme^{-\im \theta} \operatorname{det} U)^{-1/N} + \mathrm{c.c.} \right),
    \label{eq:4DchiralLagrangian_proposal}
\end{align}
where $(\det U)^{\pm 1/N}$ is a genuine local operator as the phase of $\operatorname{det} U$ is extended to have $2\pi N$ periodicity. 
We note that the large-$N$-motivated log-det vertex can be reproduced in the large-$N$ limit, so one may think that the last term is the natural finite-$N$ counterpart of the log-det vertex\footnote{We note that $(\im \ln \det U+\theta)^2$ is not a genuine local operator at finite $N$, and $(\det U)^{\pm 1/N}$ is the smallest operator consistent with the symmetry and locality. 
The $(\det U)$-type $\eta'$ mass would be unnatural when $\eta'$ is $2\pi N$ periodic because of the following features: (i)~It excludes possible smaller operators without reasoning. (ii)~Classical vacua are $N$-fold degenerate even when $m>0$ and $\theta=0$.},\footnote{It is possible to deform the last term while keeping the Witten-Veneziano relation. For example, one may consider $-\frac{1}{2}[a_1 (\rme^{-\im \theta} \operatorname{det} U)^{-1/N}+a_2( \rme^{-\im \theta} \operatorname{det} U)^{-2/N}+\mathrm{c.c.}]= -a_1 \cos\frac{\eta'-\theta}{N}-a_2 \cos\frac{2(\eta'-\theta)}{N}$, with $a_1+4 a_2=N^2 \chi_{\mathrm{YM}}$, and the $4$-point self-interaction of $\eta'$ can be arbitrarily changed including its sign. Although $a_2\sim O(\rme^{-2 S_I/N})$ is suppressed in the $2$d semiclassical regime, there is no reason to assume its smallness in the $4$d situation. We note that the following discussion is unaffected by those deformations. }.

To discuss the 't~Hooft anomaly of this theory, let us first observe the $CP$ transformation at $\theta = \pi$. 
The naive action of the $CP$ transformation is $U(\bm{x},x_4)\to U^\dagger(-\bm{x},x_4)$, 
but it changes $\left( (\rme^{-\im \pi} \operatorname{det} U)^{-1/N} + \mathrm{c.c.} \right)$ into $\left( (\rme^{\im \pi} \operatorname{det} U)^{-1/N} + \mathrm{c.c.} \right)$. 
To make it the symmetry of \eqref{eq:4DchiralLagrangian_proposal}, this must be compensated by $\operatorname{det} U \rightarrow \rme^{-2 \pi \im} \operatorname{det} U $, i.e.
\begin{equation}
    CP_{\theta=\pi}: U(\bm{x},x_4)\to \rme^{2\pi \im/N_f} U^{\dagger}(-\bm{x},x_4),
\end{equation}
and thus $\eta' \rightarrow -\eta' + 2\pi$.
Now, we remember how the $U(1)_B$ background couples to $\eta'$. 
In the presence of the $U(1)_B$ background, the $CP$ transformation of \eqref{eq:etaU1Bcoupling} leaves an extra term,
\begin{align}
    \frac{\im}{4 \pi N} \int \diff  A_B \wedge \diff A_B \in \frac{2 \pi \im}{N} \mathbb{Z}, \label{eq:lesson-would-be-anomaly}
\end{align}
which signals the potential existence of 't~Hooft anomalies, and we should now ask if this can or cannot be eliminated by local counterterms.
By considering the full flavor symmetry $U(N_f)/\mathbb{Z}_N$, possible counterterms are restricted.
In terms of the field strength $\mathcal{F}$ of $U(N_f)/\mathbb{Z}_N$ (with $\operatorname{tr}_f \mathcal{F} = N_f \diff A_B/N$), the relevant counterterm is,
\begin{align}
    \frac{\im \theta_{c}}{8 \pi^2} \int \operatorname{tr}_f \mathcal{F} \wedge \mathcal{F}.
\end{align}
We can decompose $\mathcal{F}$ into the proper $U(N_f)$ field strength $F_f$ and the $\mathbb{Z}_N$ two-form field $B$, $\mathcal{F} = F_f + B$.
This two-form background corresponds to the deviation from the standard $U(N_f)$ cocycle condition.
The $\mathbb{Z}_N$ two-form $B$ gives the fractional part of $U(1)_V$ background, $\frac{1}{N} \diff A_B =  \diff A_V + B$.
With this decomposition, the counterterm reads,
\begin{align}
    \frac{\im \theta_{c}}{8 \pi^2} \int \operatorname{tr}_f \mathcal{F} \wedge \mathcal{F} = \frac{\im \theta_{c}}{8 \pi^2} \int \left(  \operatorname{tr}_f F_f \wedge F_f + 2 \operatorname{tr}_f F_f \wedge B + N_f B \wedge B\right).
\end{align}
From the first and second terms, we require $\theta_{c} = \pi N p$ with some integer $p$.
By the $CP$ transformation, the last term yields an extra term,
\begin{align}
    -\frac{\im N N_f p}{4 \pi} \int B \wedge B = -\frac{\im (N_f p)}{4 \pi N} \int \diff  A_B \wedge \diff A_B~~(\operatorname{mod} 2 \pi \im).
\end{align}
Therefore, there must exist the integer $p$ satisfying $N_f p = 1~(\operatorname{mod} N)$ in order for the anomalous term (\ref{eq:lesson-would-be-anomaly}) to be canceled by the local counterterm.
This is possible if and only if $\operatorname{gcd}(N,N_f) = 1$.
The anomaly structure is exactly the same as that of the UV theory~\cite{Gaiotto:2017tne}.

Hence, the extension of $\eta'$ periodicity is a natural prescription for the $U(N_f)$ chiral Lagrangian to capture the global structure both not only in the $2$d compactified setup but also in $4$ dimensions. 
As we will see below, the $2 \pi N$ extended periodicity indeed plays an essential role in understanding QCD vacuum structures in the semiclassical description.

\section{\texorpdfstring{$N_f = 1$}{Nf=1} QCD}
\label{sec:NF-1}

In this and subsequent sections, we examine the QCD vacuum structure using the 2d effective description presented in Section \ref{sec:2d_EFT}.
This section is dedicated to examining one-flavor QCD, and we will consider the multi-flavor QCD ($N_f \geq 2$) in the next section (Sec.~\ref{sec:multiflavor}).

The potential (\ref{eq:one-flavor-potential}) is expressed as a superposition of the mass term, $\cos \eta'$, and the center-vortex term, $\cos \left( \frac{\eta' - \theta}{N} \right)$.
Nontrivial phenomena happen on the negative real axis (i.e. $m>0$ and $\theta = \pi$).
In short, two vacua emerge for large mass, and a unique vacuum, $\langle \eta' \rangle = \pi$, appears for small mass\footnote{The one-flavor QCD does not have the baryon-color-flavor anomaly due to $\operatorname{gcd} (N,N_f) = 1$, so the trivially gapped phase appearing for $m < m_0$ is allowed.}.
We also comment on the domain wall.

\subsection{\texorpdfstring{$CP$}{CP} breaking and restoration at \texorpdfstring{$\theta = \pi$}{theta=pi}}

\begin{figure}[t]
\centering
\includegraphics[scale=0.66]{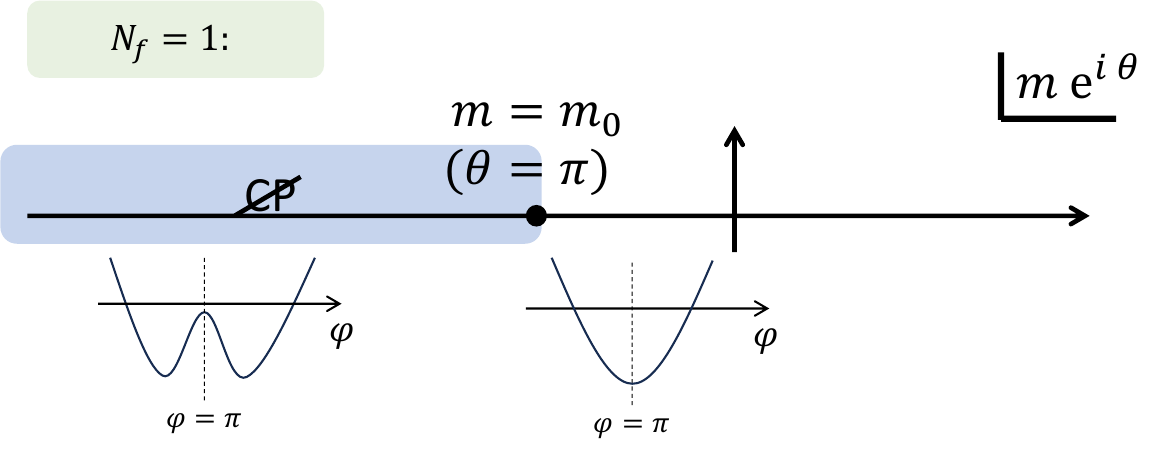}
\caption{Schematic picture of the semiclassical potential $V[\langle\eta'\rangle =\varphi]$ (\ref{eq:one-flavor-potential}) on the $m \rme^{\im \theta}$ plane. At $\theta = \pi$, i.e., on the negative real axis, the potential has two degenerate vacua for the large mass $m > m_0$. For the small mass $m < m_0$, the potential has the unique minimum at $\langle \eta' \rangle = \pi$.}
\label{fig:one-flavor-schematic}
\end{figure}

We derive the phase diagram on $m \rme^{\im \theta}$ based on the semiclassical description.
The particularly interesting parameter is $\theta = \pi$, so we focus on it.
Let us consider the small-mass and large-mass cases of the potential (\ref{eq:one-flavor-potential}). We depict these behaviors of the potential schematically in Fig.~\ref{fig:one-flavor-schematic} (which also includes the phase diagram shown in the top panel of Fig.~\ref{fig:multi_flavor_phase_diagram}).
\begin{itemize}
    \item small $m$ ($m < m_0 = \frac{2 K \rme^{-S_I/N}}{N^2 \mu}$):

    The potential has the unique minimum at $\langle \eta' \rangle = \pi$.

    \item large $m$ ($m > m_0 = \frac{2 K \rme^{-S_I/N}}{N^2 \mu}$):

    The potential has two degenerate vacua. For the large quark mass $m$, they are at $\langle \eta'\rangle \approx 0, 2 \pi$, and 
    these vacua are related by the $CP$ symmetry transformation at $\theta = \pi$, $\eta' \rightarrow - \eta' + 2\pi$.

    \item At the critical $m = m_0$, the potential $V[\eta']$ is locally flat at the origin, which indicates that $\eta'$ becomes massless.
\end{itemize}
%These observations explain the phase diagram of the one-flavor QCD, the top panel of Fig.~\ref{fig:multi_flavor_phase_diagram}.
This behavior of the semiclassical potential resembles the large-$N$ argument with the quadratic $\eta'$ mass term discussed in \cite{Rosenzweig:1979ay, Nath:1979ik, Witten:1980sp, DiVecchia:1980yfw, Gaiotto:2017tne, DiVecchia:2017xpu}.
We here use the $2\pi N$-periodic $\eta'$ field with the $\cos \left( \frac{\eta'  - \theta}{N} \right)$-type potential, and our description smoothly reproduce those previous studies in the large-$N$ limit.

In the large $m$ limit, there are $N$ vacuum candidates $\langle \eta' \rangle = 0, 2\pi, \cdots, 2 \pi (N-1)$. The $O(m^0)$ corrections splits the $N$ vacua, and the $k$-th vacuum $\langle \eta' \rangle = 2 \pi k$ ($k=0,\cdots, N-1$) has the vacuum energy:
\begin{align}
    \mathcal{E}_k(\theta) = V[\langle\eta'\rangle = 2 \pi k] = - 2 K \rme^{-S_I/N} \cos\left(\frac{2 \pi k - \theta}{N}\right) + O(m^{-1}) +\mathrm{const.}
\end{align}
This exactly recovers the semiclassical picture for the vacuum structure (\ref{eq:vac_ene_YM}) for $SU(N)$ pure Yang-Mills theory.

\subsection{Comments on domain wall}

In the $CP$-broken phase, for $m > m_0$ at $\theta = \pi$, there are two degenerate vacua. We can then consider 3d domain walls interpolating between these vacua.
In \cite{Gaiotto:2017tne}, a domain-wall phase transition has been discussed.
The main points are as follows.
Near $m_0$, the potential for $\eta'$ will be essentially the doubly-degenerate $\phi^4$ potential, leading to a trivial domain wall.
On the other hand, in the large $m$ limit, the theory is well described by the $SU(N)$ pure Yang-Mills theory, so the domain wall theory is the $SU(N)$ level-$1$ Chern-Simons theory.
Hence, there should be a domain wall transition somewhere while the bulk is smooth for $m>m_0$ at $\theta = \pi$.

In the 2d effective theory after the $T^2$ compactification, this domain wall is a kink connecting the two vacua.
However, despite the $4$d picture, the kink (or domain wall) in the $2$d effective description is smooth for $m>m_0$.
This kink does not have any nontrivial topological degrees of freedom even for large $m$.
We would like to make some comments on this ``discrepancy''.

First, the bulk adiabatic continuity only asserts the continuity in the low-energy limit, and the (bulk) massive excitations may change.
Therefore, in principle, the absence of the domain-wall phase transition does not contradict the adiabatic continuity.

Note that the domain wall becomes non-dynamical in the large mass limit $m \rightarrow \infty$.
The absence of a domain-wall phase transition is attributed to the problem that the domain wall is non-dynamical in $SU(N)$ Yang-Mills theory in our framework.
In the original 4d $SU(N)$ Yang-Mills theory, the domain wall is a dynamical object described by the $SU(N)$ level-$1$ Chern-Simons theory.
However, in our semiclassical treatment for $SU(N)$ Yang-Mills theory, the domain wall between two distinct vacuum labels is non-dynamical\footnote{As explained in \cite{Gaiotto:2017tne}, in the original 4d theory, it becomes favorable for the $\eta'$ field to ``jump over the singular point'' (which means the shift of the vacuum label) at a certain value of mass rather than to interpolate the two minima smoothly.
In the semiclassical treatment, the ``jumping over the singular point'' of the $\eta'$ potential is energetically unfavorable because the gluonic fields are almost frozen at the small $T^2$ compactification (see footnote~\ref{ftnt:2piNextension} and also the last two paragraphs of Sec.~\ref{sec:adiabatic_continuity} on the domain wall).}.

In other words, the topologically nontrivial domain wall connecting two distinct vacuum labels is decoupled due to the small $T^2$ compactification.
Therefore, the current framework is insufficient to discuss a domain wall phase transition, and we need to consider higher Kaluza-Klein modes for this purpose. 

\section{\texorpdfstring{$N_f \geq 2$}{Nf>=2} flavor-symmetric QCD}
\label{sec:multiflavor}

Let us move on to the multi-flavor QCD with degenerate quark masses.
For $N_f \geq 2$, the 2d effective description consists of one extended $2\pi N$-periodic scalar $\eta'$ and $(N_f - 1)$ $2\pi$-periodic scalars $\{ \varphi_1, \cdots ,\varphi_{N_f-1} \}$. The potential $V[\eta',\varphi_1, \cdots ,\varphi_{N_f-1}]$ for these compact scalars  is given by (\ref{eq:potential_multiflavor}).
In this section, we see that this semiclassical description explains the $CP$-breaking vacua at $\theta = \pi$ (bottom panel of Fig.~\ref{fig:multi_flavor_phase_diagram}).

\subsection{\texorpdfstring{$CP$}{CP} breaking at \texorpdfstring{$\theta = \pi$}{theta=pi}}

To study vacuum structure, we shall find minima of the potential $V[\eta',\varphi_1, \cdots ,\varphi_{N_f-1}]$.
We will put the following ansatz to simplify the computations, which is equivalent to assuming that $SU(N_f)_V$ is unbroken\footnote{In terms of the non-Abelian bosonization, this ansatz corresponds to $U \propto \textbf{1}_{N\times N}$, and it is evident that $SU(N_f)_V$ is respected.}:
\begin{align}
    \eta' &= N_f \varphi + 2 \pi k ~~~(k \in \mathbb{Z}_N) ,\notag \\
    \varphi_1 &= \cdots = \varphi_{N_f-1} = \varphi . \label{eq:symmetric_ansatz}
\end{align}
This ansatz reduces the variables to ($\rme^{\im \varphi} ,k$) $\in U(1) \times \mathbb{Z}_N$.
Let us take the branch of $\varphi$ to be $-\pi<\varphi < \pi$, and then the potential $V[\eta',\varphi_1, \cdots ,\varphi_{N_f-1}]$ becomes
\begin{align}
    V[\varphi,k] &=  - N_f m \mu \cos(\varphi) - 2 K \rme^{-S_I/N} \cos\left(\frac{N_f \varphi + 2 \pi k - \theta}{N}\right).  
\end{align}

We can observe CP-breaking doubly-degenerate vacua at $\theta = \pi$:
In the $k=0$ sector, 
\begin{align}
    V[\varphi,k=0] &=  - N_f m \mu \cos(\varphi) - 2 K \rme^{-S_I/N} \cos\left(\frac{N_f \varphi - \pi}{N}\right),
\end{align}
there is a unique minimum $\varphi = \varphi_* \in (0, \frac{\pi}{N_f})$.
This is a $CP$-breaking vacuum; the partner can be found in the $k=1$ sector
\begin{align}
    V[\varphi,k=1] &=  - N_f m \mu \cos(\varphi) - 2 K \rme^{-S_I/N} \cos\left(\frac{N_f \varphi + \pi}{N}\right),
\end{align}
which has a minimum at $\varphi = -\varphi_* $.
One can quickly check that the other sectors cannot give the minimum value.

In terms of the original variable, these vacua are
\begin{align}
    \eta' &= N_f \varphi_*,~~~\varphi_1 = \cdots = \varphi_{N_f-1} = \varphi_*,
\end{align}
($U = \rme^{i \varphi_*} \textbf{1}_{{N_f}\times {N_f}}$ in the non-Abelian bosonization), and
\begin{align}
    \eta' &= -N_f \varphi_*+2\pi,~~~\varphi_1 = \cdots = \varphi_{N_f-1} = -\varphi_*,
\end{align}
($U = \rme^{-i \varphi_*} \textbf{1}_{{N_f}\times {N_f}}$ in the non-Abelian bosonization).
Thus, the semiclassical description explains the phase diagram, or the bottom panel of Fig.~\ref{fig:multi_flavor_phase_diagram}.

\subsection{Comments on baryon-color-flavor anomalies}

The symmetry structure of $N_f$-flavor QCD is
\begin{align}
    \frac{SU(N)_{\mathrm{gauge}} \times SU(N_f)_{\mathrm{flavor}} \times U(1)_V}{\mathbb{Z}_{N} \times \mathbb{Z}_{N_f}}
\end{align}
From the quotient structure $\mathbb{Z}_{N} \times \mathbb{Z}_{N_f}$, as explored in \cite{Gaiotto:2017tne}, we can find a global inconsistency between the global symmetry and the $2 \pi$ shift of the $\theta$ angle (or anomaly with $CP$ symmetry) when $\operatorname{gcd}(N,N_f) \neq 1$. Before concluding this section, let us discuss how this anomaly manifests within the semiclassical framework.

Previously, within the ansatz, the semiclassical description consists of reduced variables $(\rme^{\im \varphi},k) \in U(1) \times \mathbb{Z}_N$. 
Notably, when $\operatorname{gcd}(N,N_f) = 1$ and only in this case, the label $k$ can be absorbed by extending $\varphi$: $(N_f \varphi + 2 \pi k) \rightarrow N_f \varphi ~~(\operatorname{mod} 2 \pi N)$. Thus, we can express the periodic scalar $\varphi \sim \varphi + 2 \pi N$ with the potential
\begin{align}
    V[\varphi] = - N_f m \mu \cos(\varphi) - 2 K \rme^{-S_I/N} \cos\left(\frac{N_f \varphi - \theta}{N}\right)
\end{align}
This expression seems quite similar to the one-flavor $N_f = 1$ case, where the scenario with a small $m$ leads to a unique vacuum. However, the term $\cos\left(\frac{N_f \varphi - \theta}{N}\right)$ exhibits $N_f$ minima unlike the one-flavor case, implying that the double degeneracy at $\theta=\pi$ persists.

For $\operatorname{gcd}(N,N_f) \neq 1$, while it is still possible to extend $\varphi$, the vacuum label of $\mathbb{Z}_{\operatorname{gcd}(N,N_f)}$ remains. This inevitability of the vacuum labels can be interpreted as a consequence of the anomaly.
This can also be understood as follows.
Recall that the minimum $\eta' = 2 \pi k$ in the large-mass limit corresponds to the $k$-th vacuum in the Yang-Mills theory.
The continuous variable $\varphi$ in the ansatz\footnote{Note that the $SU(N_f)$ symmetry forces $U = \rme^{\im \varphi} \textbf{1}$, which means the above ansatz (\ref{eq:symmetric_ansatz}) in the Abelian bosonization.} can connect the $k$-th vacuum and $(k+N_f)$-th vacuum.
If $\operatorname{gcd}(N,N_f) \neq 1$, the continuous variable cannot connect the $k=0$ vacuum and $k=1$ vacuum, which are interchanged by the $CP$ transformation at $\theta = \pi$.
In other words, if $\operatorname{gcd}(N,N_f) \neq 1$, the quark fluctuations do not mix the $k=0$ and $k=1$ branches.
The vacuum degeneracy will survive, which is a manifestation of the anomaly.

Next, let us comment on the absence of anomaly for $\operatorname{gcd}(N,N_f) = 1$.
A clear way to understand the absence of anomaly is to realize the symmetric unique gapped vacuum (trivial vacuum) by some perturbation without breaking the symmetry.
Suppose a term like
\begin{align}
    \cos\left(\frac{\varphi - \theta}{N}\right)
\end{align}
could be constructed, it is possible to restore the $CP$ symmetry (like the small-mass case of $N_f = 1$ QCD). 
Both (1) mass vertex $\rme^{\im \varphi}$ and (2) center vortex vertex $\rme^{\im \frac{ N_f \varphi - \theta}{N}}$ are consistent with symmetry, 
so the operators
\begin{align}
    (\operatorname{tr}U)^{K_1}\{(\rme^{-\im \theta}\det U)^{1/N}\}^{K_2}\sim
    (\rme^{\im \varphi})^{K_1} \left(\rme^{\im \frac{N_f \varphi - \theta}{N}}\right)^{K_2}
\end{align}
with arbitrary integers $K_1, K_2$ are allowed as symmetric perturbations.
When $\operatorname{gcd}(N,N_f) = 1$, we can choose integers $K_1$ and $K_2$ such that $N K_1 - N_f K_2 = 1$, 
which ends up with giving the $\cos\left(\frac{\varphi - K_2 \theta}{N}\right)$ term.
Therefore, if $\operatorname{gcd}(N,N_f) = 1$, we can achieve a trivial vacuum for all $\theta \sim \theta +2\pi$ by perturbation of symmetry-allowed operators, which means the absence of anomaly or global inconsistency.

However, it is important to note that unless such higher-order terms become dominant, the phase transition at $\theta = \pi$ cannot be eliminated.
In the large-mass regime, the $CP$ breaking in Yang-Mills theory persists.
In the small-mass regime, remnants of the spontaneous breakdown of the discrete chiral $(\mathbb{Z}_{N_f})_{\mathrm{chiral}}$ symmetry ($\varphi \rightarrow \varphi + \frac{2 \pi N}{N_f}$) gives two degenerate vacua (after including the mass perturbation $N_f m \mu \cos(\varphi)$) at $\theta = \pi$.
The linear combination of mass term and center vortex term, 
\begin{align}
    V[\varphi] = - N_f m \mu \cos(\varphi) - 2 K \rme^{-S_I/N} \cos\left(\frac{N_f \varphi - \theta}{N}\right),
\end{align}
necessarily gives the two degenerate vacua at $\theta = \pi$. Therefore, the $CP$-breaking at $\theta=\pi$ exists for QCD with any degenerate flavors within the semiclassical description even when the anomaly does not exist. 

\subsection{``Pion mass'' at small quark mass}

As the 2d effective theory is derived from the assumption of adiabatic continuity, it is less significant to discuss quantitative aspects.
Nevertheless, as this semiclassical description naturally bridges the chiral limit \hl{($m \rightarrow 0$)} and the quenched limit \hl{($m \rightarrow \infty$)}, it would be worthwhile to study some quantities related to the vacuum structure.
In the rest of this section, let us make some observations of the pion mass and topological susceptibility.
Here, we estimate the pion mass at a small quark mass.
Its asymptotic behavior in the massless limit could be possibly kept by adiabatic continuity.

The pion mass can be simply estimated by observing the mass for $\varphi_1, \cdots, \varphi_{N_f-1}$\footnote{The non-Abelian bosonization would be more transparent for charged pions as they become solitons in Abelian bosonization. Here, we still continue to use Abelian bosonization and restrict our attention to neutral pion, which does not change the conclusion.}.
Let us first remember that the semiclassical vacuum can be found by assuming
\begin{align}
    \eta' &= N_f \varphi + 2 \pi k ~~~(k \in \mathbb{Z}_N) \notag \\
    \varphi_1 &= \cdots = \varphi_{N_f-1} = \varphi,~~~(- \pi < \varphi < \pi)
\end{align}
with the reduced potential 
\begin{align}
    V[\varphi,k] &=  - N_f m \mu \cos(\varphi) - 2 K \rme^{-S_I/N} \cos\left(\frac{N_f \varphi + 2 \pi k - \theta}{N}\right).
\end{align}
For $- \pi < \theta < \pi$, the minimum exists at $k = 0$. The potential at $k = 0$ is,
\begin{align}
    V[\varphi,k = 0] &=  - N_f m \mu \cos(\varphi) - 2 K \rme^{-S_I/N} \cos\left(\frac{N_f \varphi - \theta}{N}\right),
\end{align}
which has a minimum between $\varphi = 0$ and $\varphi = \theta/N_f$.
When the quark mass is small, the minimum $\varphi_*$ approaches to the latter $\varphi_* \simeq \theta/N_f$.
More explicitly, we can write $\varphi_* = \theta/N_f + O(m)$.

We consider the ``pion'' fluctuations of $\varphi_1, \cdots, \varphi_{N_f-1}$, which were Nambu-Goldstone modes in the absence of quark mass.
If we look at $\varphi_j = \varphi_* + \delta \varphi_j$, the relevant terms are
\begin{align}
    - m \mu \cos(\eta' - \varphi_1 - \cdots \varphi_{N_f-1} )  - m \mu \cos(\varphi_j ) 
    &=  -  m \mu \cos(\varphi_* - \delta \varphi_j ) -  m \mu \cos(\varphi_* + \delta \varphi_j ) \notag \\
    &= -  2 m \mu \cos(\varphi_* ) \cos( \delta \varphi_j ).
\end{align}
In terms of chiral Lagrangian (or almost equivalently non-Abelian bosonization), the mass term reads,
\begin{align}
    -m \mu \left( \rme^{\im \varphi_*} \operatorname{tr} U + \mathrm{c.c.} \right),
\end{align}
where $U$ is redefined as $U \rightarrow \rme^{\im \varphi_*} U$ to denote the deviation from the minimum $\rme^{\im \varphi_*} \textbf{1}$.
Thus, the ``pion'' mass can be estimated as $m^2_\pi \sim m \mu \cos(\varphi_* )$.
For general $(N_f,|\theta|<\pi)$, it is $m^2_\pi = O(m)$ in accordance with the orthodox \hl{partially conserved axial current (PCAC)} relation.

There is an exception: If $\varphi_* = \pi/2 + O(m)$, and the pion mass shows an ``unorthodox relation'' $m^2_\pi = O(m^2)$, which has been observed in \cite{Aoki:2014moa}.
This happens if and only if $N_f = 2$, $\theta = \pi$, so
the unorthodox relation is the unique characteristic of the 2-flavor QCD.
Since the minimum $\varphi_*$ is located between $\varphi = 0$ and $\varphi = \theta/N_f$, the minimum cannot approach to $\pi/2$ unless $N_f = 2$.
Again, although the bulk adiabatic continuity does not preserve the excitations, the infrared tail, or asymptotic behavior in the massless limit, could be maintained. It is quite suggestive that our $2$d observation nicely corresponds to that of $4$d chiral Lagrangian. 
%Thus, the behavior such as the ``unorthodox relation'' $m^2_\pi = O(m^2)$ could be regarded as a nontrivial prediction.

%\red{すごい$N_f = 2$のpeculiarityのような気がしてくるが，$N_f = 2+1$ (with finite strange mass) でもunorthodox relationはでる (大事なのは$\varphi_* = \pi/2 + O(m)$なので)}

\subsection{Topological susceptibility}

\hl{
The topological susceptibility $\chi_{\mathrm{top}}$ is defined as the correlation of the topological charge density.
This quantity is often studied as a characteristic quantity that represents the fluctuation of the topological charge, e.g., see \cite{Vicari:2008jw} for a review.
In terms of the vacuum energy, it quantifies the response to changes in the $\theta$ angle: $\mathcal{E} = \frac{1}{2} \chi_{\mathrm{top}} \theta^2 + O(\theta^4)$.
}

We can obtain the topological susceptibility from the perturbation theory in $\theta$ for the vacuum energy:
\begin{align}
    \chi_{\mathrm{top}} = \frac{ N_f  m \mu \chi_{\mathrm{YM}}}{ N_f m \mu +  N_f^2 \chi_{\mathrm{YM}}}
\end{align}
where $\chi_{\mathrm{YM}}$ is the topological susceptibility of the pure Yang-Mills theory,
\begin{align}
    \chi_{\mathrm{YM}} = \frac{2 K \rme^{-S_I/N}}{N^2}.
\end{align}
This expression interpolates between the chiral limit $\chi_{\mathrm{top}} = 0$ and the quenched limit $\chi_{\mathrm{top}} = \chi_{\mathrm{YM}}$.\footnote{This functional form is identical to the result from the large-$N$ $\eta'$ mass term of Refs.~\cite{Witten:1980sp, DiVecchia:1980yfw, Sakai:2004cn}, which was used in Ref.~\cite{Mameda:2014cxa}. 
We note, however, that the coincidence is somewhat trivial as this calculation only uses the $\theta$-derivatives at $\theta=0$ and does not utilize the global structure. Therefore, the consequence does not depend upon whether the contribution from center-vortex-induced gluonic effects is represented by a simple cosine function, its $1/N$ fractionalization, or a quadratic function. }

\section{Dashen phase}
\label{sec:Dashen_phase}

In the context of the strong $CP$ problem, there is an interest in the physics that occurs when varying the mass of the up quark in $(1+1)$-flavor and $(1+1+1)$-flavor QCD. For example, although it is no longer thought to be a plausible scenario~\cite{Nelson:2001bhq}, the fine-tuning problem of the theta angle would not exist if the up quark were massless. Furthermore, intriguing phase structures are expected on the $(m_u, m_d)$-plane.
It is anticipated that, when the mass of the up quark takes a large negative value beyond a certain point, a spontaneously $CP$-broken phase, known as the Dashen phase \cite{Dashen:1970et}, would appear. On the phase boundary of this Dashen phase, the neutral pion becomes massless.

In the following of this section, we set $\theta=0$, and $m_u$, $m_d$ are varied on both the positive and negative values following the convention of Refs.~\cite{Creutz:1995wf, Creutz:2003xc,Creutz:2003xu, Creutz:2013xfa, Aoki:2014moa}. 
When one of $m_u$, $m_d$ is negative, i.e. $m_u m_d<0$, it corresponds to $\theta=\pi$ with $m_u, m_d>0$ in the convention of previous sections. 

In Refs.~\cite{Creutz:1995wf, Creutz:2003xc,Creutz:2003xu, Creutz:2013xfa, Aoki:2014moa}, the phase diagram on the $(m_u, m_d)$-plane has been investigated using the chiral Lagrangian (without and with $\eta'$)\footnote{What we call $\eta'$ corresponds to $\eta=u\bar{u}+d\bar{d}$ in the genuine $2$-flavor case, but let us simply call the flavor-singlet component as $\eta'$ for any flavors.}. It has been asserted that nothing special happens at $m_u = 0$ except at $m_d = 0$, and the existence of the Dashen phase has been confirmed with the appearance of a massless pion at the phase boundaries.
In these references, however, a simple $U(N_f)$-valued meson field with a $\operatorname{det} U$-type $\eta'$ mass term was employed. As seen before (Section~\ref{sec:lesson}), the periodicity extension of $\eta'$ is important to capture the global perspective correctly, so let us re-examine the discussions about the Dashen phase from the semiclassical framework.

\subsection{Dashen phase of \texorpdfstring{$(1+1)$}{(1+1)}-flavor QCD}

First, we shall see the Dashen phase in the semiclassical description of non-degenerate two-flavor QCD.
The semiclassical 2d effective theory consists of $2\pi N$-periodic $\eta'$ and $2 \pi$-periodic $\varphi$ with the following potential (\ref{eq:potential_multiflavor}),
\begin{align}
    V[\eta',\varphi] &= - m
_u \mu \cos(\eta' - \varphi) - m_d \mu \cos (\varphi) - 2 K \rme^{-S_I/N} \cos\left(\frac{\eta' - \theta}{N}\right).
\end{align}
Here, we focus on the phase diagram on $(m_u,m_d)$ plane at $\theta = 0$. 
The potential has the $CP$ invariance: $(\eta',\varphi) \rightarrow (-\eta',-\varphi)$.
An obvious candidate of the vacuum is the $CP$ symmetric one $\langle \eta'\rangle = \langle \varphi \rangle = 0$, which is indeed the ground state for $m_u > 0$ and $m_d >0$.
For $m_u < 0$ and $m_d < 0$, the ground state is another $CP$ symmetric vacuum $\langle \eta'\rangle = 0,~\langle \varphi\rangle = \pi$.

The phase boundary between the $CP$-symmetric and $CP$-broken phases can be analytically obtained by considering the stability of these $CP$-symmetric vacua $\eta' = \varphi = 0$ and $\eta' = 0,~\varphi = \pi$.
For the former vacuum $\eta' = \varphi = 0$, the mass matrix has zero eigenvalue when 
\begin{align}
    m_u m_d + \Delta (m_u+m_d) = 0,
    \label{eq:pioncondensate1}
\end{align}
where we have defined $\Delta := \frac{2 K \rme^{-S_I/N}}{N^2 \mu}$.
This determines the phase boundary of the trivial phase $\eta' = \varphi = 0$, and it is a rectangular hyperbola with asymptotes $m_u = -\Delta$ and $m_d = -\Delta$.
Similarly, the phase boundary of the latter vacuum $\eta' = 0,~\varphi = \pi$ is located on
\begin{align}
    m_u m_d - \Delta (m_u+m_d) = 0,
    \label{eq:pioncondensate2}
\end{align}
which is a rectangular hyperbola with asymptotes $m_u = \Delta$ and $m_d = \Delta$.

\begin{figure}[t]
\centering
\includegraphics[scale=0.7]{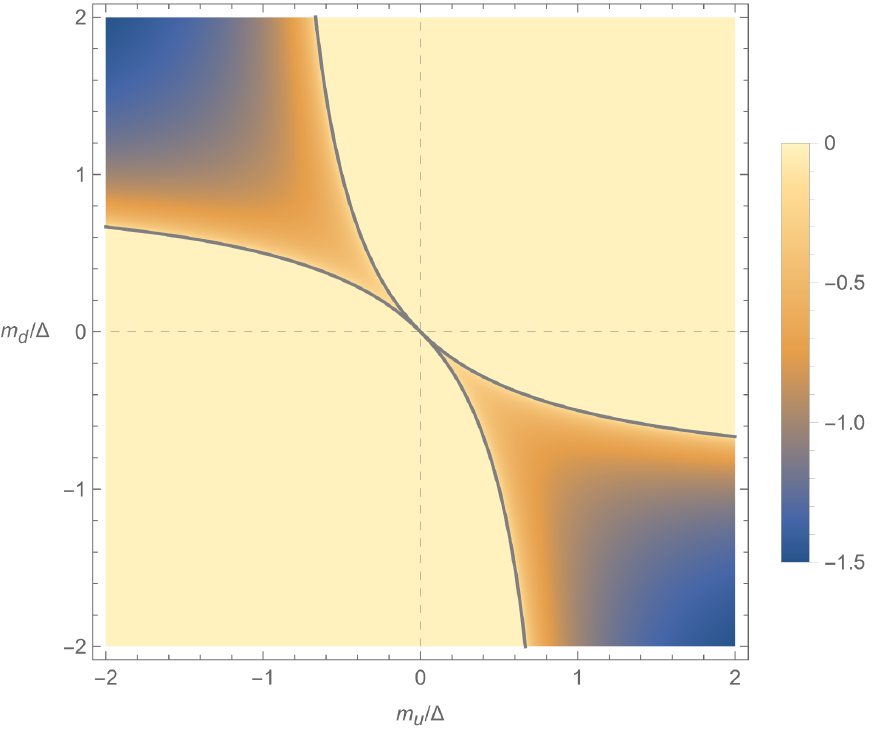}
\caption{
Plot of vacuum expectation value $\langle \eta'\rangle$ on the $(m_u/\Delta,m_d/\Delta)$ plane with $\Delta = \frac{2 K \rme^{-S_I/N}}{N^2 \mu}$ and $\theta=0$. %The values of $\eta'$ are computed for the case $\Delta=1$ as demonstration. 
The $CP$ symmetry is preserved in the trivial phases ($\eta' = 0$), whereas the $CP$ symmetry is spontaneously broken in the $\eta' \neq 0$ phase, which is the Dashen phase. In this figure, we show the $\eta'<0$ configuration of the two $CP$-broken vacua.
The phase boundaries are analytically determined by $m_u m_d + \Delta (m_u+m_d) = 0$ and $m_u m_d - \Delta (m_u+m_d) = 0$, as explained in the main text.}
\label{fig:AC_2flavor}
\end{figure}

The phase boundaries, together with the numerical plot of minimum $\eta'$, are depicted in Fig.~\ref{fig:AC_2flavor}.
The intermediate region, where $\eta' \neq 0$, is the $CP$-broken ``pion''-condensed Dashen phase.
This result supports the main features expected in the previous studies:
There is the Dashen phase for $m_u m_d <0$, the massless mode (``pion'') appears on its phase boundary, and the phase is trivial on the line $m_u = 0$ except for the origin $m_u = m_d = 0$. 
These phase boundaries given by \eqref{eq:pioncondensate1} and \eqref{eq:pioncondensate2} are exactly identical to those found in \cite{Aoki:2014moa} (up to some parametrization), where $U(2)$ chiral Lagrangian with the $(\operatorname{det} U)$-type $\eta'$ mass is employed.
Since the above discussion only relies on the stability of the trivial vacua, the phase boundary does not depend on the global structure.

The difference between our result and previous ones appears when the global structure is involved.
For example, a $CP$ symmetric phase (``phase C'' in Ref.~\cite{Aoki:2014moa}) appears for large $(m_u, -m_d)$, i.e., in the limit of $m_u \rightarrow \infty$ and $m_d \rightarrow -\infty$, in their model.
This $CP$ restoration needs to be understood as an artifact of the $U(2)$ chiral Lagrangian: QCD in the large-$(m_u, -m_d)$ limit reduces to the pure Yang-Mills theory at $\theta = \pi$, where the spontaneous breakdown of $CP$ symmetry is expected.\footnote{Also in Ref.~\cite{Aoki:2014moa}, the authors made a comment in one of footnotes that ``phase C'' disappears if one use the large-$N$-type log-det $\eta'$ mass instead, and we agree on this point. However, the $CP$ breaking at $\theta=\pi$ is required by anomaly matching, so it must occur for any choice of the symmetry-preserving potentials, including the $(\det U)$-type $\eta'$ mass. 
Our interpretation is that the target space has to be $2\pi N$-extended from $U(N_f)$, and then the vacuum configuration of their ``phase C'' at large-$(m_u, -m_d)$ also violates the $CP$ symmetry.%, while it is separated from another $CP$-broken phase at small mass with $m_u=-m_d$.
%Thus, our proposal to correctly interpret their phase diagram is that $CP$ is always broken when $m_u=-m_d$ as requested by anomaly but $CP$-broken phases at small and large masses with $m_u=-m_d$ are separated by accidental phase transitions. This accident occurs due to the fine-tuning that excludes the lower-dimensional operator $(\det U)^{1/N}$ while the higher-dimensional operator $(\det U)$ is included. In this sense, the use of $\det U$-type $\eta'$ mass does not fit the naturalness criterion. 
\label{footnote:periodicity}}
Similarly, in the decoupling limit $m_u \rightarrow + \infty$, the $m_d$-dependence should be identical to that of $N_f = 1$ QCD (Fig.~\ref{fig:one-flavor-schematic}). 
We can see that the periodicity extension of $\eta'$ plays an important role when the gluonic multi-branch vacuum structure is significant.

\subsection{Dashen phase of \texorpdfstring{$(1+1+1)$}{(1+1+1)}-flavor QCD}

The above phase diagram \hl{is} affected by the strange quark.
Next, let us consider the phase diagram of $N_f = 1 + 1 +1$ QCD on the $(m_u,m_d)$ plane with fixed strange mass $m_s>0$ and $\theta=0$.
For large $m_s$, the qualitative feature of the phase diagram is unchanged as it is essentially $(1+1)$-flavor QCD.
A nontrivial phenomenon happens if the strange mass is below the critical mass of the 1-flavor QCD (see Fig.~\ref{fig:one-flavor-schematic}).
In this case, the $CP$ restored phase appears in the limit $(m_u,m_d) \to (\infty,-\infty)$ and $(m_u,m_d) \to (-\infty,\infty)$.

Let us study how the phase diagram is changed by the strange quark.
The semiclassical potential (at $\theta = 0$) is,
\begin{align}
    V[\eta' , \varphi_1, \varphi_2] &= - m
_s \mu \cos(\eta' - \varphi_1 - \varphi_2) - m_u \mu \cos (\varphi_1)- m_d \mu \cos (\varphi_2) - 2 K \rme^{-S_I/N} \cos\frac{\eta'}{N}.
\end{align}
Immediately, we can observe that there are $CP$-symmetric phases in some particular regions:
\begin{itemize}
    \item For $m_u > 0$ and $m_d > 0$, the vacuum is $(\eta' , \varphi_1, \varphi_2) = (0,0,0)$. Similarly, for $m_u < 0$ and $m_d < 0$, the vacuum is $(\eta' , \varphi_1, \varphi_2) = (0,\pi,\pi)$.
    \item New $CP$-restored phases appear if $m_s$ is below the critical mass, $m_s < \Delta = \frac{2 K \rme^{-S_I/N}}{N^2 \mu}$.
    Then, for $-m_u \gg m_s$ and  $m_d \gg m_s$, the vacuum is $(\eta' , \varphi_1, \varphi_2) = (0,\pi,0)$.
    Similarly, for $m_u \gg m_s$ and  $-m_d \gg m_s$, the vacuum is $(\eta' , \varphi_1, \varphi_2) = (0,0,\pi)$.
\end{itemize}

To determine phase boundaries, we now consider the stability of these $CP$-symmetric minima.
By a simple computation of the mass matrix, the phase boundary of each trivial phase can be obtained as follows:
For the former phases, we have
\begin{align}
    (m_s + \Delta) m_u m_d + (m_s \Delta) (m_u + m_d) = 0~~~\mathrm{for~}(\eta' , \varphi_1, \varphi_2) = (0,0,0)~\mathrm{phase}, \label{eq:phase_bdy_A}\\
    (m_s + \Delta) m_u m_d - (m_s \Delta) (m_u + m_d) = 0~~~\mathrm{for~}(\eta' , \varphi_1, \varphi_2) = (0,\pi,\pi)~\mathrm{phase}. \label{eq:phase_bdy_B}
\end{align}
The asymptotes are $m_u = \pm \frac{m_s \Delta}{m_s + \Delta}$ and $m_d = \pm \frac{m_s \Delta}{m_s + \Delta}$.
For the latter phases, the phase boundaries suggested by the Hessian determinants are,
\begin{align}
     - (-m_s + \Delta) m_u m_d + (-m_s \Delta) (m_u - m_d) = 0~~~\mathrm{for~}(\eta' , \varphi_1, \varphi_2) = (0,0,\pi)~\mathrm{phase}, \label{eq:phase_bdy_C} \\
    - (-m_s + \Delta) m_u m_d + (-m_s \Delta) (-m_u + m_d) = 0~~~\mathrm{for~}(\eta' , \varphi_1, \varphi_2) = (0,\pi,0)~\mathrm{phase}, \label{eq:phase_bdy_D}. 
\end{align}
These extrema are unstable for $m_s \geq \Delta$\footnote{For $m_s \geq \Delta$, the Hessian determinant is positive but it means two unstable modes.}.
For subcritical strange mass $m_s < \Delta$, the phase boundaries of these phases are determined by the above hyperbolas with asymptotes $m_u = \pm \frac{m_s \Delta}{\Delta - m_s}$ and $m_d = \pm \frac{m_s \Delta}{\Delta - m_s}$.

For supercritical strange mass $m_s \geq \Delta$, the phase diagram is qualitatively the same as the 2-flavor one shown in Fig.~\ref{fig:AC_2flavor}.
The asymptotes of the phase boundaries are affected to be $m_u = \pm \frac{m_s \Delta}{m_s + \Delta}$ and $m_d = \pm \frac{m_s \Delta}{m_s + \Delta}$, which were $m_u = \pm \Delta$ and $m_d = \pm \Delta$ in the 2-flavor case.

\begin{figure}[t]
\centering
\includegraphics[scale=0.7]{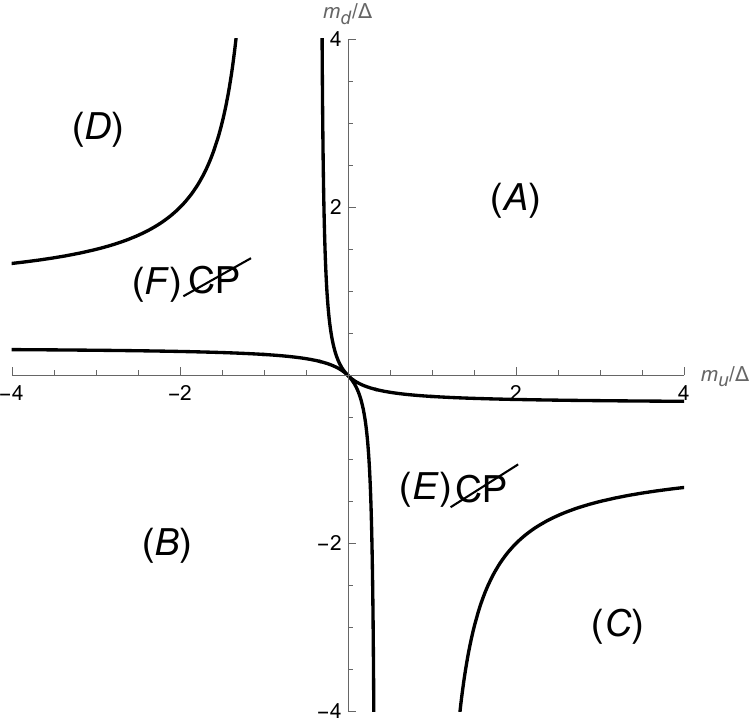}
\caption{
The phase diagram on the $(m_u,m_d)$ plane for the $N_f = 1+1+1$ QCD at a subcritical strange mass $m_s = \Delta/2 (<\Delta)$.
The four regions (A) -- (D) are $CP$-symmetric trivial phases, and the other regions (E) and (F) are $CP$-broken Dashen phases.
}
\label{fig:3flavor_phase_diagram_mumd}
\end{figure}

The phase diagram substantially changes for subcritical regime $m_s < \Delta$.
There exists a $CP$-restored phase for large $(m_u,-m_d)$ and $(-m_u,m_d)$.
For example, a phase diagram at $m_s = \Delta/2$ is plotted in Fig.~\ref{fig:3flavor_phase_diagram_mumd}.
There are four CP symmetric trivial phases: 
\begin{enumerate}
\renewcommand{\labelenumi}{(\Alph{enumi})}
    \item $(\eta' , \varphi_1, \varphi_2) = (0,0,0)$. 
    
    This phase includes the first quadrant $m_u>0$ and $m_d > 0$. The phase boundary is given by (\ref{eq:phase_bdy_A}).

    \item $(\eta' , \varphi_1, \varphi_2) = (0,\pi,\pi)$

     This phase includes the third quadrant $m_u < 0$ and $m_d < 0$.
     The phase boundary is given by (\ref{eq:phase_bdy_B}).

     \item $(\eta' , \varphi_1, \varphi_2) = (0,0,\pi)$.

     This phase appears in the lower-right region including $m_u \gg m_s$ and $-m_d  \gg m_s$.
     The phase boundary is given by (\ref{eq:phase_bdy_C}).

      \item $(\eta' , \varphi_1, \varphi_2) = (0,\pi,0)$.

This phase appears in the upper-left region including $-m_u \gg m_s$ and $m_d  \gg m_s$.
     The phase boundary is given by (\ref{eq:phase_bdy_D}).
\end{enumerate}
The remaining regions (E) and (F) are the $CP$-broken Dashen phases.

This phase diagram bears a striking resemblance to the one depicted in \cite{Creutz:2003xu}. In that work, a similar phase diagram on the $(m_u, m_d)$ plane was drawn from the chiral Lagrangian without $\eta'$. Several comments are in order regarding this observation:

\begin{itemize}
    \item In the limit where the $\eta'$ meson decouples, $\Delta \rightarrow \infty$, the phase boundaries coincide with those of the chiral Lagrangian, as expected.
    
    \item A difference can be found in the decoupling limit of $m_u$ or $m_d$. For instance, in the limit $m_d \rightarrow +\infty$, the chiral Lagrangian without $\eta'$ suggests that the Dashen phase converges to a point. In contrast, our results indicate a Dashen phase in the range $- \frac{m_s \Delta}{\Delta - m_s } < m_u < - \frac{m_s \Delta}{\Delta + m_s }$. Notably, as $m_s$ approaches $\Delta$, this width increases, and for $m_s \geq \Delta$, the $CP$-broken phase covers the semi-infinite region $m_u < - \frac{m_s \Delta}{\Delta + m_s }$.
    
    \item In \cite{Creutz:2003xu}, the $U(3)$ chiral Lagrangian with $(\operatorname{det} U)$-type $\eta'$ mass term was also examined, concluding that it does not qualitatively differ from the chiral Lagrangian. 
    This incorporation of the $\eta'$ meson gives the qualitatively same phase diagram as Fig.~\ref{fig:3flavor_phase_diagram_mumd},
    However, even with a heavy strange mass $m_s > \Delta$, the phases (C) and (D) artificially persist in the $U(3)$ chiral Lagrangian\footnote{For instance, instead of phase (C), the $U(3)$ chiral Lagrangian yields an artificial trivial phase for $m_s > \Delta$: At this vacuum, the chiral field $U$ is,
    \begin{align*}
        U = \begin{pmatrix}
1 & 0 & 0\\
0 & -1 & 0 \\
0 & 0 & 1
\end{pmatrix}
    \end{align*}
    which is equivalent to $\varphi_1 = 0, \varphi_2 = \pi, \eta' = \pi$ in our parameterization.
    In our model, this configuration is not $CP$ invariant: $(\varphi_1, \varphi_2, \eta') = (0,\pi,\pi)$ and $(\varphi_1, \varphi_2, \eta') = (0,\pi,-\pi)$ are discriminated due to the periodicity extension 
    (See also footnote~\ref{footnote:periodicity}). 
    %As noted in the footnote~\ref{footnote:periodicity}, the large and small $m_u=-m_d$ regions are both $CP$-broken but separated by the accidental phase transition when the $(\det U)$-type $\eta'$ mass is used, which is caused by the fine-tuning of excluding the lower-dimensional operator, $(\det U)^{\pm 1/N}$
    }. 
    As previously discussed, this artificial phase is inconsistent with the phase diagram of the one-flavor QCD at $\theta = \pi$ (Fig.~\ref{fig:one-flavor-schematic}) in the decoupling limit of $m_u \rightarrow +\infty$ and $m_d\rightarrow -\infty$.
    The use of $(2 \pi N)$-periodic $\eta'$ meson improves the consistency with global aspects.

\end{itemize}

In conclusion, the semiclassical framework successfully explains the main characteristics of the Dashen phase that are anticipated in prior studies.
In comparison to the conventional chiral Lagrangian, one of its advantages is the inherent inclusion of the gluonic multi-branch structure leading to the periodicity extension of $\eta'$ as noted in Sec.~\ref{sec:lesson}.

\section{Conclusion}

In this work, we have explored the QCD vacuum structure, employing a novel semiclassical approach on $\mathbb{R}^2 \times T^2$ with the 't Hooft and $U(1)_B$ fluxes, proposed in \cite{Tanizaki:2022ngt}. 
Our main finding is that the simple semiclassical 2d description can successfully explain the well-known vacuum structures of QCD (Fig.~\ref{fig:multi_flavor_phase_diagram}).

A key implication of our approach is about the incorporation of the $\eta'$ meson into the chiral Lagrangian (Section \ref{sec:lesson}). 
In our semiclassical framework, the periodicity of $\eta'$ is extended from $2\pi$ to $2 \pi N$. 
This periodicity extension is explicitly derived in $2$d effective description, and then we argue that this extension of the $\eta'$ should be true also in the original $4$d setup. 
Assuming this extension, we can align the $4$d chiral Lagrangian including the $\eta'$ with various global structures:
For example, we can reproduce the discrete anomalies, as well as the vacuum structure of the quenched limit, which could not be explained by the naive $U(N_f)$ chiral Lagrangian with the $2\pi$ periodic $\eta'$. 
We also propose the $4$d chiral Lagrangian~\eqref{eq:4DchiralLagrangian_proposal} consistent with the $2\pi N$-extended $\eta'$ field. 

Furthermore, our application of this framework to the Dashen phases has yielded refined insights (Sections \ref{sec:NF-1}-\ref{sec:Dashen_phase}).
We have confirmed that our semiclassical framework effectively accounts for the key features of the Dashen phase as predicted in earlier research. 
Compared to approaches from the conventional chiral Lagrangian, a notable benefit of this framework is its natural incorporation of the gluonic multi-branch structure through the extension of periodicity in $\eta'$.
Our result refines the understanding of this point.

We would like to re-emphasize that our 2d effective description is microscopically obtained, not derived phenomenologically. Although the adiabatic continuity conjecture is a key assumption to apply it for the $4$d vacuum strcuture, the 2d effective description itself is derived from the original QCD Lagrangian (at small $T^2$ compactification).
Additionally, our results have not only reinforced the widely accepted vacuum structure of QCD, in turn, but also have provided (indirect) evidence supporting the validity of the adiabatic continuity conjecture.

There are several possible avenues for future studies.
A notable advantage of our semiclassical 2d effective description, in contrast to the traditional chiral Lagrangian, lies in its improved treatment of the $\eta'$ meson. 
Therefore, it would be interesting to explore physics where the global structure of the $\eta'$ meson plays a crucial role.
Additionally, it would be worthwhile to further investigate other models based on the semiclassical framework, including different-rank bifundamental QCD and supersymmetric theories.
For a more fundamental direction, an intriguing (but difficult) prospect is to understand if there exists a smooth and explicit connection between monopole and center-vortex semiclassics, thereby deepening our understanding of the confinement mechanism.

\acknowledgments

The authors thank Sinya Aoki, Aleksey Cherman, and Hiromasa Watanabe for useful discussions and comments on the draft. 
The work of Y. T. was supported by Japan Society for the Promotion of Science (JSPS) KAKENHI Grant numbers, 22H01218, and by Center for Gravitational Physics and Quantum Information (CGPQI) at Yukawa Institute for Theoretical Physics.
Y.~H. was supported by JSPS Research Fellowship for Young Scientists Grant No.~23KJ1161

\appendix

\section{Bosonized Schwinger model: an exercise for extending periodicity}
\label{sec:extendedperiodicity}

As an example of extending the periodicity of a compact scalar done in Sec.~\ref{sec:QCD_center_vortex}, we consider the bosonized charge-1 Schwinger model, namely 2d QED. 
After the bosonization, the action reads,
\begin{align}
    S[a,\phi] = \int \frac{1}{2 e^2} |\diff a|^2 + \frac{\im \theta}{2 \pi} \diff a  + \frac{\im }{2 \pi}   a \wedge \diff\phi + \frac{1}{8\pi} |\diff \phi|^2, 
\end{align}
where $a$ is the $U(1)$ gauge field, $\phi$ is the $2 \pi$ periodic bosonized variable, and $|f|^2 := f \wedge \star f$.
Note that the vector-like current $\Bar{\psi}\gamma_\mu \psi$ of the fermion is translated to the topological current $\frac{\im }{2 \pi}  \diff \phi$ in the bosonization. 

A common manipulation is to integrate out the $U(1)$ gauge field, and we get the mass term $\frac{e^2}{8 \pi^2} (\phi + \theta)^2$.
Naively, this appears inconsistent with the periodicity, and let us carefully consider the integration of the $U(1)$ gauge field step-by-step (see also \cite{Armoni:2018bga, Misumi:2019dwq, Tanizaki:2019rbk, Cherman:2020cvw, Cherman:2021nox, Cherman:2022ecu}).

The integration over the $U(1)$ gauge field consists of the following summations
\begin{itemize}
    \item $U(1)$ gauge bundles (Chern class), 
    \item constant holonomies, and
    \item local fluctuations (globally-defined $1$-form field),
\end{itemize}
and similarly, a $2\pi$ periodic compact scalar can be represented by a set of
\begin{itemize}
    \item ``local fluctuations'': a globally-defined $\Tilde{\phi} \in \mathbb{R}$
    \item ``topological sectors'': $c/2\pi \in H^1 (M_2;\mathbb{Z})$.
\end{itemize}

Here, the constant holonomy only appears in the topological coupling
\begin{align}
    \frac{\im }{2 \pi}   a \wedge \diff\phi,
\end{align}
and its integration\footnote{If the summation over the constant holonomy were over $\mathbb{Z}_N$, the constrain would be $c/(2\pi N) \in H^1 (M_2;\mathbb{Z})$.
This happens in the main text, Sec.~\ref{sec:QCD_center_vortex}.} constraints the nontrivial winding: $c/2\pi = 0$. 
For example, on $T^2$ with the size $L$, the $2\pi$-periodic field $\phi$ may have the globally winding configuration such as $\phi(x_1,x_2)=2\pi (n_1 \frac{x_1}{L}+n_2\frac{x_2}{L})$ with $n_1,n_2\in \mathbb{Z}$, but the holonomy integration sets $n_1,n_2=0$. 
This enables us to regard $\phi$ as a globally-defined $\mathbb{R}$-valued field.

To integrate out the $U(1)$ gauge field, we represent the field strength $\diff a$ as a 2-form field $f$ with the Dirac quantization condition $\int \frac{f}{2 \pi} \in \mathbb{Z}$. 
To be more precise, the field strength needs to be written as $\diff a = \diff \Tilde{a} + 2 \pi m$, where $\Tilde{a}$ is a globally-defined 1-form representing the local fluctuations and $m \in H^2 (M_2;\mathbb{Z})$ is the Chern class specifying the topological sector.
The integration over the $U(1)$ gauge field can be expressed as,
\begin{align}
    \int \mathcal{D}a ~ (\cdots)=   \sum_{m \in H^2 (M_2;\mathbb{Z})} \int \mathcal{D}\Tilde{a} \left. ~ (\cdots) \right|_{\diff a = \diff \Tilde{a} + 2 \pi m}.
\end{align}
Let us rewrite it into the another expression. 
The $U(1)$ field strength $f = \diff a$ can also be characterized by the closedness $\diff f = 0$ and the Dirac quantization $\int \frac{f}{2 \pi} \in \mathbb{Z}$.
The former condition is trivial on the 2d manifold, so we can represent the integration over the $U(1)$ gauge field as the integration over the 2-form field with the quantization constraint:
\begin{align}
    \int \mathcal{D}a ~ (\cdots) &=   \int \mathcal{D} f \left. ~ (\cdots) \right|_{\diff a = f} \times \delta \left( \int \frac{f}{2\pi} \in \mathbb{Z} \right) \notag \\
    &= \sum_{n \in \mathbb{Z}} \int \mathcal{D} f ~ \rme^{\im n \int f} \left. ~ (\cdots) \right|_{\diff a = f}.
\end{align}
One may regard that this manipulation is the $2$d Abelian duality between the $1$-form and $(-1)$-form gauge fields, $\star \diff a \leftrightarrow n$.

Now, the action becomes
\begin{align}
    S = \int \frac{1}{2 e^2} \left| f + \frac{\im e^2}{2 \pi} (\phi + \theta) \star 1  \right|^2   + \frac{e^2}{8 \pi^2} (\phi + \theta)^2  \star 1 + \frac{1}{8\pi}  |\diff \phi|^2,
\end{align}
and we can integrate out the $U(1)$ gauge field as follows:
\begin{align}
    \sum_{n \in \mathbb{Z}} & \int \mathcal{D} f  ~ \rme^{\im n \int f} \rme^{- \int \frac{1}{2 e^2} \left| f + \frac{\im e^2}{2 \pi} (\phi + \theta) \star 1  \right|^2} \notag \\ 
    &= \sum_{n \in \mathbb{Z}} \rme^{ n \int \frac{e^2}{2 \pi} (\phi + \theta) \star 1 } \int \mathcal{D} f ~ \rme^{\im n \int f} \rme^{- \int \frac{1}{2 e^2} \left| f  \right|^2} \notag \\
    &= \left(  \int \mathcal{D} f ~ \rme^{- \int \frac{1}{2 e^2} \left| f  \right|^2} \right) \sum_{n \in \mathbb{Z}} \rme^{ \int \left[ -\frac{e^2}{2} n^2 + \frac{ne^2}{2 \pi} (\phi + \theta) \right] \star 1 }. 
\end{align}
Therefore, we can evaluate the partition function as
\begin{align}
    Z = \int \mathcal{D}\phi \int \mathcal{D}a ~ \rme^{-S[a,\phi]} \propto \int_{[\diff \phi] = 0 } \mathcal{D}\phi \sum_{n \in \mathbb{Z}} \rme^{- \int \frac{e^2}{8 \pi^2} (\phi + \theta - 2 \pi n)^2  \star 1 - \frac{1}{8\pi}  |\diff \phi|^2 }.
\end{align}
We note that $\phi$ in the last expression is subject to the constraint $[\diff \phi] = 0 \in H^1(M_2;\mathbb{Z})$, which means that $\phi$ cannot have a nontrivial global winding, and this constraint appears as a result of the integration over the holonomies.

From the constraint $[\diff \phi] = 0$, we can lift $\phi$ to a $\mathbb{R}$-valued field $\Tilde{\phi}$, 
but there are (infinitely) many lifts $\{ \Tilde{\phi} + 2 \pi n \}_{n \in \mathbb{Z}}$ that gives the identical $\phi \in (\mathbb{R}/2\pi \mathbb{Z})$.
Thus, the integration over a $\mathbb{R}$-valued field can be written as,
\begin{align}
    \int \mathcal{D}\Tilde{\phi}~ (\cdots) = \int_{[\diff \phi] = 0 } \mathcal{D}\phi \sum_{n \in \mathbb{Z}}~ \left. (\cdots) \right|_{ \Tilde{\phi} = \phi - 2 \pi n  },
\end{align}
where $\phi$ in the substitution ``$\Tilde{\phi} = \phi - 2 \pi n$'' means an arbitrary lift of $\phi$ to a $\mathbb{R}$-valued field.
The sum over $n \in \mathbb{Z}$ is a summation of all smooth lifts of $(\mathbb{R}/2\pi \mathbb{Z})$-valued field $\phi$.
As in the main text, the integration over $U(1)$ gauge field, which magnetically couples to the compact scalar, extends the periodicity: from a $(\mathbb{R}/2\pi \mathbb{Z})$-valued field $\phi$ to a $\mathbb{R}$-valued field $\Tilde{\phi}$.
We then end up with the effective theory for the periodicity-extended $\mathbb{R}$-valued field $\Tilde{\phi}$,
\begin{align}
    Z = \int \mathcal{D}\phi \int \mathcal{D}a ~ \rme^{-S[a,\phi]} \propto  \int \mathcal{D}\Tilde{\phi}~ \rme^{- \int \frac{e^2}{8 \pi^2} (\Tilde{\phi} + \theta)^2  \star 1 - \frac{1}{8\pi}  |\diff \Tilde{\phi}|^2 }.
\end{align}
Essentially, the same phenomenon happens in the 2d effective theory for QCD on $\mathbb{R}^2 \times T^2$.

\begin{table}[t]
    \centering
\begin{tabular}{ |l|c|c| } 
 \hline
Model & Schwinger model & 2d EFT of QCD on $\mathbb{R}^2 \times T^2$ \\
 \hline
\underline{\textbf{Elements}} & & \\
Gauged symmetry of compact boson & $U(1)_{\mathrm{mag}}$ & residual $(\mathbb{Z}_N)_{\mathrm{mag}}$\\
Magnetic object & $U(1)$ flux & $\mathbb{Z}_N$ center vortex \\
 \hline
 \multicolumn{2}{|l|}{\underline{\textbf{Consequences by integrating out the gauge field}}} &  \\
Periodicity extension & $\mathbb{R}$-valued $\Tilde{\phi}$ & $\eta' \sim \eta' + 2 \pi N$ \\
Induced mass term & $\frac{e^2}{8 \pi} (\Tilde{\phi} + \theta)^2 $ & $- 2 K \rme^{-S_I/N} \cos \left( \frac{\eta' - \theta}{N} \right)$ \\
 \hline
\end{tabular}
    \caption{Comparison between the Schwinger model and the $T^2$-flux compactified (1-flavor) QCD discussed in the main text.}
    \label{tab:comparison}
\end{table}

Let us turn on the mass perturbation $-\mu m \cos(\phi)$ and briefly discuss the vacuum structure at $\theta = \pi$.
The argument is quite parallel to that of the one-flavor QCD presented in Sec.~\ref{sec:NF-1}.
A comparison between these two models is summarized in Table \ref{tab:comparison}.
The qualitative structure can be readily extracted from the potential:
\begin{align}
    V(\Tilde{\phi}) = -\mu m \cos(\phi) + \frac{e^2}{8 \pi^2} (\Tilde{\phi} + \theta)^2.
\end{align}
When the mass term is weak $\mu m < \frac{e^2}{4 \pi}$, we have the unique vacuum $\Tilde{\phi} = - \pi$.
On the other hand, for large mass  $\mu m > \frac{e^2}{4 \pi}$, the extrema $\Tilde{\phi} = - \pi$ is not a minimum, and two minima appear.
We obtain the similar phase diagram as Fig.~\ref{fig:one-flavor-schematic} for the 1-flavor Schwinger model.

\bibliographystyle{JHEP}
\bibliography{./QFT.bib, ./adds.bib} 

\end{document}